%% file: block.tex
\documentclass[journal, letterpaper]{IEEEtran}
\usepackage[left=0.53in, right=0.53in, top=0.8in, bottom=0.45in]{geometry}
\usepackage{cite}

\usepackage{hyperref}%
\hypersetup{%
  breaklinks=true,%
  colorlinks=false,%
  linkcolor=blue,%
  citecolor=blue,%
  filecolor=blue,%
  urlcolor=blue
  }%

\usepackage[labelfont=footnotesize, textfont=footnotesize,labelformat=simple,labelsep=period, aboveskip=3pt,belowskip=0pt]{caption}
\usepackage[labelfont=footnotesize,textfont=footnotesize, labelformat=simple,labelsep=period, aboveskip=-10pt,belowskip=5pt]{subcaption}

\usepackage{color}%
\usepackage{bbm}%
\usepackage{stmaryrd}
\usepackage{mathtools} 
\usepackage{amssymb, amsopn}%
\usepackage[ntheorem, framemethod=TikZ]{mdframed} 
\usepackage{framed}  
\usepackage[amsmath, hyperref, framed]{ntheorem}%
\usepackage{bm}%
\usepackage{upgreek}
\usepackage{mathdots}
\mathtoolsset{showonlyrefs, showmanualtags}

\AtBeginDocument{\let\citep\cite}
\AtBeginDocument{\let\citet\cite}

\usepackage{ifthen}%
\usepackage{xkeyval}%
\usepackage{xfor}%
\usepackage{amsgen}%
\usepackage{etoolbox}%
\usepackage{datatool-base}%

\usepackage[%
  xindy,%
  style=long,%
  toc=true,%
  nonumberlist,%
  acronym,%
  acronymlists={block_abbreviations},%
  hyperfirst=true
  ]{glossaries}%

\input{block_abbreviations}%

\makeglossaries%

\definecolor{plotgreen}{rgb}{0.2509804, 0.5607843, 0.2823529}
\definecolor{plotmagenta}{rgb}{0.7764706, 0.1254902, 0.8117647}
\definecolor{plotblue}{rgb}{0.1764706, 0.1607843, 0.9607843}

\usepackage[inline]{enumitem}

\newlist{myenumerate}{enumerate}{3} 
\setlist[myenumerate,1]{label=\arabic*), ref=\arabic*}
\setlist[myenumerate,2]{label=\roman*), ref=\roman*}
\setlist[myenumerate,3]{label=\alph*), ref=\alph*}

\newlist{myenumerateAlg}{enumerate}{3} 
\setlist[myenumerateAlg]{itemsep=1ex, topsep=1ex}
\setlist[myenumerateAlg,1]{label=\arabic*), ref=\arabic*, leftmargin=*, itemsep=0.5ex, topsep=0.5ex}
\setlist[myenumerateAlg,2]{label=\roman*), ref=\roman*, labelindent=\parindent, topsep=0.5ex, itemsep=0.5ex}
\setlist[myenumerateAlg,3]{label=\alph*), ref=\alph*, labelindent=\parindent, topsep=0.5ex, itemsep=0.5ex}

\newlist{mydescription}{enumerate}{3} 

\qedsymbol{\ensuremath{\Box}}

\theoremstyle{plain}%

\theoremheaderfont{\indent\itshape\rmfamily} 

\theorembodyfont{\normalfont\itshape}%
\theoremseparator{:}

\newtheorem{proposition}{Proposition}%
\newtheorem{lemma}{Lemma}%
\newtheorem{assumption}{Assumption}%
\newtheorem{definition}{Definition}%
\newtheorem{remark}{Remark}%
\newtheorem{example}{Example}%

\theorembodyfont{\normalfont}%

\newmdtheoremenv[
 ntheorem=true,
 skipbelow = .6\baselineskip plus 1ex minus 1ex,
 skipabove = .6\baselineskip plus 1ex minus 1ex,
 innerleftmargin = 0pt,
 innerrightmargin = 0pt,
 leftline = false,
 rightline = false,
 needspace = 5ex 
]{framedAlgorithm}{Algorithm}

\theoremstyle{nonumberplain}%
\theoremheaderfont{\indent\rmfamily\itshape}%
\theorembodyfont{\upshape \mdseries}%
\theoremsymbol{\ensuremath{_\Box}}
\theoremsymbol{\ensuremath{_\square}}
\newtheorem{proof}{Proof}%

\input{block_mathematics}%

\newcommand{\IE}{i.e.\ }
\newcommand{\EG}{e.g.\ }
\newcommand{\WRT}{w.r.t.\ }
\newcommand{\RHS}{r.h.s.\ }

\glsresetall

\begin{document}

\title{Approximate Smoothing and Parameter Estimation\\in High-Dimensional State-Space Models}

\author{Axel~Finke, \and Sumeetpal~S.~Singh\\\vspace{-0.05cm}
\thanks{This work was supported by the Engineering and Physical Sciences Research Council [grant number EP/K020153/1].}
\thanks{A.~Finke is with the Department of Statistical Science, University College London, UK (e-mail: a.finke@ucl.ac.uk)}
\thanks{S.~S.~Singh is with the Department of Engineering, University of Cambridge, UK and the Alan Turing Institute, London, UK (e-mail: sss40@cam.ac.uk)}
\thanks{This article has been accepted for publication in a future issue of IEEE Transactions on Signal Processing, but has not been fully edited. Content may change prior to final publication. Citation information: \href{https://doi.org/10.1109/TSP.2017.2733504}{DOI 10.1109/TSP.2017.2733504}}
}



\IEEEpubid{
  \begin{minipage}{\textwidth}\ \\[20pt]
    \centering
    1053-587X (c) 2017 IEEE. Personal use is permitted, but republication/redistribution requires IEEE permission. See \href{http://www.ieee.org/publications_standards/publications/rights/index.html}{http://www.ieee.org/publications\_standards/publications/rights/index.html} for more information.
  \end{minipage}
} 


\maketitle

\begin{abstract}
   \noindent{}We present approximate algorithms for performing smoothing in a class of high-dimensional state-space models via \glsdesc{SMC} methods (`particle filters'). In high dimensions, a prohibitively large number of Monte Carlo samples (`particles'), growing exponentially in the dimension of the state space, is usually required to obtain a useful smoother. Employing blocking approximations, we exploit the spatial ergodicity properties of the model to circumvent this curse of dimensionality. We thus obtain approximate smoothers that can be computed recursively in time and parallel in space. First, we show that the bias of our blocked smoother is bounded uniformly in the time horizon and in the model dimension. We then approximate the blocked smoother with particles and derive the asymptotic variance of idealised versions of our blocked particle smoother to show that variance is no longer adversely effected by the dimension of the model. Finally, we employ our method to successfully perform maximum-likelihood estimation via stochastic gradient-ascent and stochastic \glsdesc{EM} algorithms in a 100-dimensional state-space model.
\end{abstract}


%
\IEEEpeerreviewmaketitle

\section{Introduction}
\glsresetall

\IEEEPARstart{A}{lgorithms} known as \gls{SMC} methods or \glspl{PF} are nowadays commonly used to perform inference in state-space models \citep{cappe2005inference, doucet2011tutorial} and more generally \citep{delmoral2006sequential}. In order to infer certain `static' model parameters, \gls{SMC} methods are often employed within other algorithms, \IE within \gls{MCMC} or other \gls{SMC} approaches for Bayesian inference \citep{andrieu2010particle, chopin2013smc2} or within stochastic gradient-ascent and \gls{EM} algorithms for Frequentist inference \citep{poyiadjis2011particle}. In both cases, it is imperative to control the error of \gls{SMC} approximations of the filter and smoother. Unfortunately, the number of Monte Carlo samples (`particles') typically needs to scale exponentially in the dimension of the state space in order to control these errors \citep{bengtsson2008curse}. This `curse of dimensionality' quickly leads to a prohibitive computational cost in higher dimensions. 

In order to circumvent the curse of dimensionality in the context of \emph{filtering}, so called `blocking approximations' have been introduced in the literature. For finite state-space dynamic Bayesian networks, blocked filter approximations (termed \emph{Boyen--Koller} algorithm) were proposed by \citet{boyen1998tractable, boyen1999exploiting}. \gls{SMC} approximations of the Boyen--Koller algorithm were first considered in \citet{ng2002factored}. For models on general state spaces \citet{brandao2006subspace, besada2009parallel} used similar algorithms. \citet{rebeschini2015local} termed one such algorithm \glsdesc{BPF} and showed that it permits approximations of filter marginals which are bounded uniformly in time and model dimension as long as the model is sufficiently ergodic in both time and space. Other attempts at reducing the state-dimension by introducing additional intermediate \gls{SMC} steps were made in \citet{maccormick2000probabilistic, wuthrich2015coordinate, beskos2017stable}. However, for problems without a very specific conditional-independence structure, these strategies induce models which are no longer Markov which may cause difficulties in an \gls{SMC} context. For state-space models with Gaussian-mixture transitions, the \emph{iterated auxiliary \glsdesc{PF}} introduced by \citet{guarniero2016iterated} can make Bayesian inference via particle \gls{MCMC} methods \citep{andrieu2010particle} feasible in high dimensions.

The problem of performing \emph{smoothing} for high-dimensional state-space models has received much less attention thus far. For \emph{small finite} state spaces, \citet{koller1999approximate} attempted a simple approximate forward--backward recursion and more sophisticated blocked smoothing algorithms are developed by \citet[personal communication]{alaluhtalaINPREPfactorized} independently of the present work. However, for \emph{general} (\EG large discrete or even continuous) state spaces, efficient smoothing algorithms in high dimensions are still lacking.

\IEEEpubidadjcol
The goal of this work is thus to exploit blocking strategies to devise approximate versions of particle smoothing algorithms known as \emph{\glsdesc{BS}} \citep{godsill2004monte} and \emph{\glsdesc{FS}} \citep{delmoral2010forward} for the canonical class of high-dimensional \glsdescplural{HMM} from \citet{rebeschini2015local} and to provide theoretical guarantees for these methods. Our methodological contributions, mainly contained in Section~\ref{sec:methodology}, are as follows. 
\begin{itemize}
  \item In Subsection~\ref{subsec:motivation}, we show that existing smoothing algorithms break down in high dimensions in the sense that the asymptotic variance grows exponentially in the model dimension -- even if we make the favourable assumption that the filter errors are dimension-independent 
  (Proposition~\ref{prop:exploding_asymptotic_variance}). This unequivocally justifies the need for new dimensionally stable particle smoothers.

  \item In Subsection~\ref{subsec:proposed_algorithm}, we introduce novel blocked particle smoothing algorithms and a bias-reduction technique.

  \item In Subsection~\ref{subsec:theory}, we prove a uniform (in both time and model dimension) bound on the asymptotic variance of our estimator (Proposition~\ref{prop:asymptotic_variance}) and a similarly uniform bound on the asymptotic bias (Proposition~\ref{prop:asymptotic_bias}). 

\end{itemize} 
In Section~\ref{sec:simulations}, we empirically illustrate that 
our algorithm induces local errors which are bounded in the model dimension. Finally, we successfully perform maximum-likelihood estimation via stochastic gradient-ascent and stochastic \gls{EM} algorithms in a $100$-dimensional state-space model.

\section{Standard Particle Filtering and Smoothing}
\label{sec:standard_particle_filtering_and_smoothing}
\glsreset{PF}

In this section, we review standard particle methodology for performing filtering or smoothing in state-space models.

\subsection{State-Space Models}
\label{subsec:standard_ssm}
A (homogeneous) state-space model is a stochastic process $(X_t, Y_t)_{t \in \naturals}$ on a space $\spaceX \times \spaceY$ with the following properties. The process $(X_t)_{t \in \naturals}$ is a Markov chain on $\spaceX$ with initial distribution $\Init(\diff x_1) = \init(x_1)\refMeasX(\diff x_1)$ and transitions $\trans(x_{t-1},x_t)\refMeasX(\diff x_t)$. Here, $\refMeasX$ denotes the reference measure on $\spaceX$ with respect to which $\init\colon \spaceX \to (0,\infty)$ and $\trans \colon \spaceX \times \spaceX \to (0,\infty)$ are densities. The chain $(X_t)_{t \in \naturals}$ is not directly observed. At each time~$t$, we instead observe the value $y_t$ of the random variate $Y_t$ whose law, conditional on the `state' sequence $X_{1:t} \coloneqq (X_1, \dotsc, X_t)$ taking the values $x_{1:t} \in \spaceX^t$, is $\obs(x_t, y_t)\refMeasY(\diff y_t)$. Here, $\refMeasY$ denotes the reference measure on $\spaceY$ with respect to which $\obs \colon \spaceX \times \spaceY \to (0,\infty)$ is a transition density. Throughout this work, we will refer to $\modelDim \coloneqq \card \spaceX$ as the \emph{model dimension}. Finally, all densities may depend on some model parameter $\theta$ but for simplicity, we suppress $\theta$ from the notation wherever possible.

\subsubsection{Filtering}
In many applications, \EG when tracking objects in real time, we are interested in approximating integrals with respect to the \emph{filter} at time~$t$, $\Filt_t$. This is the distribution of $X_t$ given all the observations recorded up to time~$t$, \IE $\Filt_t(\diff x_t) = \Prob(X_t \in \diff x_t| Y_{1:t} = y_{1:t})$.

\subsubsection{Smoothing}
The \emph{joint smoothing distribution} $\smooth_t$ is the posterior distribution of the states $X_{1:t}$ given the data $y_{1:t}$, \IE $\smooth_t(\diff x_{1:t}) \coloneqq \Prob(X_{1:t} \in \diff x_{1:t}| Y_{1:t} = y_{1:t})$. It is given by $\smooth_t(\diff x_{1:t}) \propto \uSmooth_t(\diff x_{1:t})$, where
\begin{align}
  \!\!\!\!\uSmooth_t(\diff x_{1:t}) = \Init(\diff x_1) \obs(x_1, y_1) \smashoperator{\prod_{s=2}^t} \trans(x_{s-1}, x_s) \obs(x_s, y_s) \refMeasX(\diff x_s).\!\!\!\!
\end{align}

\subsubsection{Goal}
In this work, we seek to efficiently approximate integrals of certain additive functions $\smoothFun_t\colon \spaceX^t \to \reals$ (see Subsection~\ref{subsec:high-dimensional_smoothing}) \WRT the joint smoothing distribution, \IE
\begin{equation}
  \smoothFunIntegral_t \coloneqq \smooth_t(\smoothFun_t) \coloneqq \E[\smoothFun_t(X_{1:t})], \quad \text{for $X_{1:t} \sim \smooth_t$},
\end{equation}
for the canonical class of high-dimensional state-space models introduced in \citep{rebeschini2015local} (and detailed in Subsection~\ref{subsec:high-dimensional_ssm} of this work). First, however, we review existing methods conventionally employed for this purpose. We will sometimes refer to these as \emph{standard} particle filters and smoothers to distinguish them from the \emph{blocked} particle filters and smoothers reviewed and introduced in Sections~\ref{sec:blocked_pf} and \ref{sec:methodology}, respectively.

\subsection{Standard Particle Filtering}

We begin by reviewing (standard) \glspl{PF}. 
Let $\Prop_t(z, \diff x) \coloneqq \prop_t(z, x) \refMeasX(\diff x)$ be a Markov transition kernel, where $\prop_t\colon \spaceX \times \spaceX \to (0,\infty)$ is some suitable transition density with respect to the reference measure $\refMeasX$. All these quantities may also depend on $y_{1:t}$ but we suppress this dependence, for simplicity. Finally, define  
\begin{align}                                                                                                                                                                                                                                                                                                                                                                         
  G_1(x)      &\coloneqq \frac{\init(x)\obs(x, y_1)}{\prop_1(x)},\\
  G_{t}(z, x) &\coloneqq \frac{\trans(z, x)\obs(x, y_t)}{\prop_t(z, x)}, \quad \text{for $t > 1$.}
\end{align}

Algorithm~\ref{alg:standard_pf} summarises a \gls{PF}. Here, $\dCat(p^{1:\nParticles})$ is the categorical distribution with some vector of probabilities $p^{1:\nParticles}$ and we use the convention that actions prescribed for the $n$th particle are to be performed \emph{conditionally independently} for all $n \in \{1, \dotsc, N\}$, where $\nParticles$ is the number of particles.

\begin{flushleft}
\begin{framedAlgorithm}[particle filter] \label{alg:standard_pf}~
\begin{flushleft}
\begin{myenumerateAlg}
 \item At time~$1$,
  \begin{myenumerateAlg} 
   \item sample $\smash{X_{1}^n \sim \Prop_{1}}$,
   \item set $\smash{w_1^n \coloneqq G_1(X_1^n)}$ and $\smash{W_1^n \coloneqq w_1^n / \sum_{k=1}^\nParticles w_1^k}$.
  \end{myenumerateAlg}

 \item\label{alg:standard_pf:stept} At Step~$t$, $t > 1$, 
  \begin{myenumerateAlg}
    \item\label{alg:standard_pf:resampling} sample $\smash{A_{t-1}^{n} \sim \dCat(W_{t-1}^{1:\nParticles}})$; write $\smash{\widebar{X}_{t-1}^n \coloneqq X_{t-1}^{A_{t-1}^n}}$,
    \item sample $\smash{X_{t}^n \sim \Prop_{t}(\widebar{X}_{t-1}^n, \ccdot)}$; 
    \item set $\smash{w_t^n \coloneqq G_t(\widebar{X}_{t-1}^n, X_t^n)}$ and $\smash{W_t^n \coloneqq w_t^n / \sum_{k=1}^\nParticles w_t^k}$.
   \end{myenumerateAlg}
 \end{myenumerateAlg}
\end{flushleft}
\end{framedAlgorithm}
\end{flushleft}

If $\prop_{t} = \trans$ then Algorithm~\ref{alg:standard_pf} is often termed \emph{bootstrap} \gls{PF}. For any test function $\testFun\colon \spaceX \to \reals$, we may approximate $\Filt_{t}(\testFun)$ using the weighted sample $(X_{t}^n, W_t^n)_{n \leq \nParticles}$. That is, after the $t$th step, we may estimate $\Filt_t(\testFun)$ by
\begin{equation}
  \Filt^\nParticles_{t,\mathsc{\pf}}(\testFun) 
  \coloneqq \sum_{\smash{n = 1}}^\nParticles W_t^n \testFun(X_t^n). \label{eq:standard_pf_filter_approximation}
\end{equation}
Throughout this work, we assume that the cost of sampling each particle $X_t^n$ and evaluating each weight $w_t^n$ grows linearly in $\modelDim$. The complexity of the \gls{PF} is then $\bo(\modelDim \nParticles)$ per time step.

\subsection{Standard Particle Smoothing}
\label{subsec:standard_particle_smoothing}

We now review algorithms which have been proposed to approximate expectations with respect to the joint smoothing distribution, \IE integrals $\smoothFunIntegral_\nTimeSteps = \smooth_\nTimeSteps(\smoothFun_\nTimeSteps)$, for some $\nTimeSteps \in \naturals$.

\subsubsection{Backward Recursion}

It is well known \citep[Corollary~3.3.8]{cappe2005inference} that the joint smoothing distribution can be written as
\begin{equation}
 \smooth_\nTimeSteps(\diff x_{1:\nTimeSteps}) = \Filt_\nTimeSteps(\diff x_\nTimeSteps) \smashoperator{\prod_{t=1}^{T-1}} \back_{ \Filt_t}(x_{t+1}, \diff x_t), \label{eq:standard_backward_recursion}
\end{equation}
where the \emph{backward kernels} $\back_{\nu} \colon \spaceX \times \Borel(\spaceX) \to [0,1]$ ($\Borel(\spaceX)$ is the Borel $\sigma$-algebra and $\nu$ some probability measure on $\spaceX$) are defined by
\begin{align}
 \back_{ \nu}(x, \diff z)
 = \frac{\trans(z, x) \nu(\diff z) }{\int_\spaceX \trans(u, x) \nu(\diff u) }.
\end{align}

\subsubsection{Particle Approximation}

Unfortunately, the filters $\Filt_t$ and hence the backward kernels $\back_{\Filt_t}(x, \diff z)$ in \eqref{eq:standard_backward_recursion} are usually intractable unless the model is linear and Gaussian or unless the model dimension $\modelDim = \card \spaceX$ is sufficiently small. To circumvent this intractability, standard particle smoothers \citep{godsill2004monte,delmoral2010forward} replace the filters in \eqref{eq:standard_backward_recursion} by Monte Carlo approximations. More precisely, let $(X_t^n, W_t^n)_{n \leq \nParticles}$ be a weighted sample (\EG obtained from the \gls{PF} in Algorithm~\ref{alg:standard_pf}) such that 
  $\Filt^\nParticles_t\coloneqq \sum_{n=1}^\nParticles W_t^n \delta_{X_t^n}$, 
approximates $\Filt_t$. Here, $\delta_x$ denotes the point mass at $x$. We then replace $\back_{\Filt_t}(x, \diff z)$ in \eqref{eq:standard_backward_recursion} by
\begin{align}
  \back_{ \Filt^\nParticles_{t}}(x, \diff z)
  & = \sum_{n=1}^\nParticles \frac{W_t^n \trans(X_t^n, x)}{\sum_{k = 1}^\nParticles W_t^k \trans(X_t^k, x)} \delta_{X_t^n}(\diff z). \label{eq:backward_kernel}
\end{align}
Conditional on $(X_t^n, W_t^n)_{n \in \{1,\dotsc,\nParticles\}}$, the computational complexity of evaluating $\smash{\back_{\Filt^\nParticles_{t}}(x, \ccdot)}$ is $\bo(\nParticles \modelDim)$.

Replacing $\smash{\back_{\Filt_t}(x, \diff z)}$ by $\smash{\back_{ \Filt^\nParticles_{t}}(x, \diff z)}$ in \eqref{eq:standard_backward_recursion} then induces the following approximation of $\smooth_\nTimeSteps(\smoothFun_\nTimeSteps)$:
\begin{gather}
  \smooth^\nParticles_{\nTimeSteps,\fs}(\smoothFun_\nTimeSteps)
  \coloneqq \int_{\smash{\spaceX^\nTimeSteps}} \smoothFun_\nTimeSteps(x_{1:\nTimeSteps}) \Filt_{\nTimeSteps}^\nParticles(\diff x_\nTimeSteps) \smashoperator{\prod_{\smash{t=1}}^{\smash{\nTimeSteps-1}}} \back_{\Filt_t^\nParticles}(x_{t+1}, \diff x_t)\\[-0.6ex]
  = \smashoperator{\sum_{n_{1:\nTimeSteps} \in \{1,\dotsc,\nParticles\}^\nTimeSteps}} \smoothFun_\nTimeSteps(X_{1:\nTimeSteps}^{n_{1:\nTimeSteps}}) W_\nTimeSteps^{n_\nTimeSteps} \prod_{t = 1}^{\smash{\nTimeSteps-1}} \! \back_{\Filt^\nParticles_{t}}(X_{t+1}^{n_{t+1}}, \{X_t^{n_t}\}). \label{eq:standard_forward_smoothing_approximation}
\end{gather}

\subsubsection{Forward Smoothing}
The computational cost of summing over $N^\nTimeSteps$ terms in \eqref{eq:standard_forward_smoothing_approximation} is normally prohibitive. However, if $\smoothFun_\nTimeSteps$ is \emph{additive in time} in the sense that there are functions $\testFun_1 \colon \spaceX \to \reals$ and $\testFun_t \colon \spaceX^2 \to \reals$ (where $\testFun_t$ may depend on $y_t$) such that
\begin{align}
  \smoothFun_t(x_{1:t}) 
  & = \testFun_1(x_1) + \sum_{s = 2}^t \testFun_s(x_{s-1},x_s), \label{eq:additive_in_time}
\end{align}
then the computational cost of evaluating this estimate can be brought down to $\bo(\nParticles^2 \nTimeSteps \modelDim)$. The resulting algorithm is called (standard) \emph{\gls{FS}} and was introduced by \citet{delmoral2010forward, delmoral2010backward} who also derived finite-sample error bounds as well as a central limit theorem (see also \citep{douc2011sequential}). Algorithm~\ref{alg:standard_fs} outlines the idea; we use the convention that any action prescribed for \emph{some} $n$ is to be performed conditionally independently for \emph{all} $n \in\{1,\dotsc, \nParticles\}$. Note that Algorithm~\ref{alg:standard_fs} can be implemented online, \IE $\alpha_t^n$ can already be determined at Step~$t$ of the \gls{PF}. 

\begin{flushleft}
\begin{framedAlgorithm}[forward smoothing] \label{alg:standard_fs}~
\begin{flushleft}
\begin{myenumerateAlg}

  \item \label{alg:standard_fs:recursion} Set $\alpha_1^n \coloneqq \testFun_1(X_1^n)$. For $t > 1$, set
   \begin{equation}
     \alpha_t^n \coloneqq \sum_{m = 1}^{\smash{\nParticles}}
   \back_{\Filt^\nParticles_{t-1}}(X_t^{n}, \{X_{t-1}^m\})
   [\alpha_{t-1}^m + \testFun_t(X_{t-1}^m, X_t^n)].
   \end{equation}
  
 \item Approximate $\smoothFunIntegral_t$ by $\smash{\smoothFunIntegral^\nParticles_t \coloneqq \smooth^\nParticles_{t,\fs}(\smoothFun_t) = \sum_{n = 1}^\nParticles W_t^n \alpha_t^n}$.
 \end{myenumerateAlg}
\end{flushleft}
\end{framedAlgorithm}
\end{flushleft}

\subsubsection{Backward Sampling}
To circumvent the $\bo(\nParticles^2)$ computational complexity of \gls{FS}, we may instead estimate $\smooth_\nTimeSteps(\smoothFun_\nTimeSteps)$ using a simple Monte Carlo approximation based on $\nBackwardPaths < \nParticles$ sample points drawn conditionally independently from $\smash{\smooth^\nParticles_{\nTimeSteps,\fs}}$. More precisely, the algorithm samples $M$ particle paths $\widetilde{X}_{1:\nTimeSteps}^m$ in the reverse-time direction according to the kernel from \eqref{eq:backward_kernel}.
This gives the (standard) \emph{\gls{BS}} \citep{godsill2004monte} approximation\footnote{Note that additivity in time of the test function is not needed for \gls{BS}.}
\begin{equation}
  \smooth^\nParticles_{\nTimeSteps,\bs}(\smoothFun_\nTimeSteps)
  \coloneqq
  \frac{1}{\nBackwardPaths}\sum_{m = 1}^\nBackwardPaths \smoothFun_\nTimeSteps(\widetilde{X}_{1:\nTimeSteps}^m). \label{eq:ffbs_self-normalised_estimate}
\end{equation}
Algorithm~\ref{alg:standard_bs} outlines the method. Here, we use the convention that any action prescribed for \emph{some} $m$ is to be performed conditionally independently for \emph{all} $m \in \{1,\dotsc, \nBackwardPaths\}$. 

\begin{flushleft}
\begin{framedAlgorithm}[backward sampling] \label{alg:standard_bs}~
\begin{flushleft}
\begin{myenumerateAlg}
 \item Sample $\smash{\widetilde{X}_\nTimeSteps^m \sim \Filt^\nParticles_{\nTimeSteps}}$ and $\smash{\widetilde{X}_t^m \sim \back_{ \Filt^\nParticles_{t}}(\widetilde{X}_{t+1}^m, \ccdot)}$, for $t < \nTimeSteps$. 
 \item Approximate $\smoothFunIntegral_\nTimeSteps$ by $\smoothFunIntegral^\nParticles_\nTimeSteps \coloneqq \smooth^\nParticles_{\nTimeSteps,\bs}(\smoothFun_\nTimeSteps)$.
 \end{myenumerateAlg}
\end{flushleft}
\end{framedAlgorithm}
\end{flushleft}

The computational complexity of Algorithm~\ref{alg:standard_bs} is $\bo(\nBackwardPaths \nParticles \nTimeSteps \modelDim)$. However, $\smooth^\nParticles_{\nTimeSteps,\bs}(\smoothFun_\nTimeSteps)$ normally has a larger variance than $\smooth^\nParticles_{\nTimeSteps,\fs}(\smoothFun_\nTimeSteps)$ since \gls{FS} can be seen as a Rao--Blackwellisation of the \gls{BS} approximation, \IE since
\begin{align}
 \smooth^\nParticles_{\nTimeSteps,\fs}(\smoothFun_\nTimeSteps) 
 & = \E\bigl[ \smooth^\nParticles_{\nTimeSteps,\bs}(\smoothFun_\nTimeSteps)\big| w_{1:\nTimeSteps}^{1:N}, X_{1:\nTimeSteps}^{1:\nParticles} \bigr]. \label{eq:fs_as_rao-blackwellised_bs}
\end{align}

As proposed in \citep{douc2011sequential}, the $\bo(\nBackwardPaths\nParticles)$-complexity of \gls{BS} can be reduced to $\bo(\nParticles)$ using an accept-reject step which circumvents the need for evaluating the denominator in \eqref{eq:backward_kernel}, and \citet{olssonEfficient} developed an online implementation around this idea called \gls{PaRIS}. However, the accept-reject step typically requires $\bo(\eul^\modelDim)$ \emph{proposed} samples to obtain a single \emph{accepted} sample. The overall complexity of \gls{PaRIS} or of the backward sampler from \citep{douc2011sequential} is therefore $\bo(\nParticles T \eul^\modelDim)$.


\section{Blocked Particle Filtering}
\label{sec:blocked_pf}

In this section, we describe the canonical class of high-dimensional state-space models for which we will compute the smoother in the next section. The same model was used in \citet{rebeschini2015local} to analyse their blocked filtering algorithm which is also reviewed.

\subsection{Class of High-dimensional State-Space Models}
\label{subsec:high-dimensional_ssm}

The state-space model $(X_t, Y_t)_{t \in \naturals}$ from Subsection~\ref{subsec:standard_ssm} is now developed into a high-dimensional model as follows. We assume that the state space $\spaceX = \prod_{v \in \vertexSet} \spaceX_v$ is endowed with a graph $\graphSet \coloneqq (\vertexSet, \edgeSet)$ where vertices $v \in \vertexSet$ index the components of the state vector and edges $e \in \edgeSet$ define the spatial correlation structure. The latent states $\smash{X_t \coloneqq (X_{t,v})_{v \in \vertexSet}}$ are then $\modelDim$-dimensional, with $V = \card \spaceX = \card \vertexSet$ again being the model dimension. For each component $X_{t,v}$, taking a value $x_{t,v} \in \spaceX_v$, we obtain an observation $Y_{t,v}$, taking a value $y_{t,v} \in \spaceY_v$. Thus, $Y_t \coloneqq (Y_{t,v})_{v \in \vertexSet}$ 
takes a value in $\spaceY = \prod_{v \in \vertexSet} \spaceY_v$. Finally, for all $v \in \vertexSet$, we let $\refMeasX_v$ and $\refMeasY_v$ be reference measures on $\spaceX_v$ and $\spaceY_v$, respectively, with $\smash{\refMeasX = \prod_{v \in \vertexSet}\refMeasX_v}$ and $\smash{\refMeasY = \prod_{v \in \vertexSet}\refMeasY_v}$. We assume that the densities satisfy the following properties.

\begin{figure}[ht]
  \centering
  \includegraphics[trim=0cm 0.1cm 0cm 0.4cm]{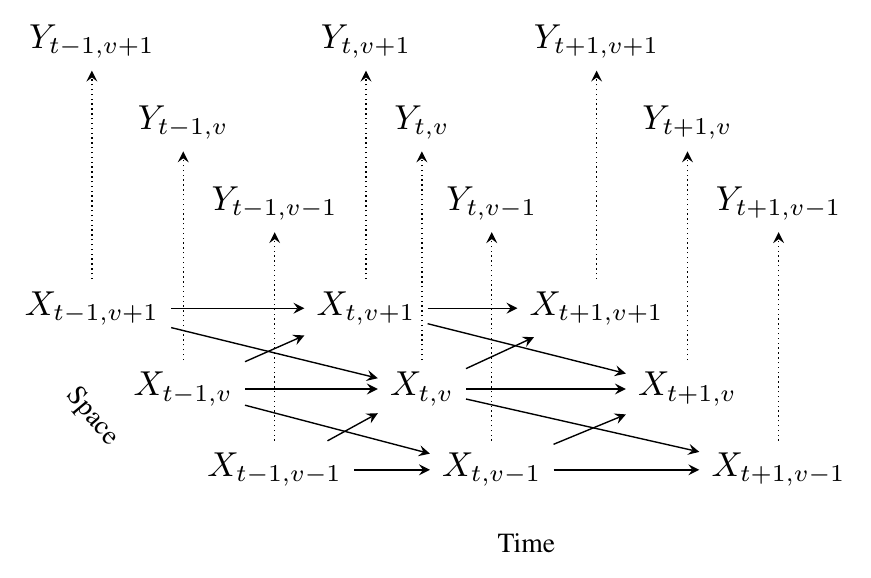}
  \caption{Sketch of the state-space model considered in this work. In this example, $\neighSize =1$, \IE $\neigh(v) = \{v-1, v, v+1\}$. A similar figure can be found in \citet{rebeschini2015local}.}
  \label{fig:ssm}
\end{figure}

\begin{myenumerate}[label=\roman*), ref=\roman*]
  \item \label{enum:model_property:1} The initial density factorises as $\init(x) = \prod_{v \in \vertexSet} \init_v(x_v)$, where $\init_v\colon \spaceX_v \to (0,\infty)$ is a density \WRT $\refMeasX_v$.
  
  \item \label{enum:model_property:2} The transition densities $p$ and $g$ factorise as
  \begin{equation}
    \trans(z, x) = \prod_{v \in \vertexSet} \trans_v(z, x_v) \;\; \text{and} \;\; \obs(x, y) = \prod_{v \in \vertexSet} \obs_v(x_v, y_v),
  \end{equation}
  where $\trans_v\colon \spaceX \times \spaceX_v \to (0,\infty)$ and $\obs_v\colon \spaceX_v \times \spaceY_v \to (0,\infty)$ are transition densities \WRT $\refMeasX_v$ and $\refMeasY_v$, respectively.
  

  \item Let $\neigh_\neighSize(v)$ be the $\neighSize$-neighbourhood of the vertex $v$, \IE
  \begin{equation}
    \neigh_\neighSize(v) \coloneqq \{u \in \vertexSet \mid \metric(u,v) \leq \neighSize\}, 
  \end{equation}
  where $\metric(u,v)$ is the length of the shortest path between vertices $u$ and $v$, and $\neighSize > 0$. The parameter $\neighSize$ is fixed throughout this work and we write $\neigh(v) \coloneqq \neigh_\neighSize(v)$. We assume that $\neighSize$ governs the spatial correlation of the model in the sense that $\trans_v(z, x_v) = \trans_v(z', x_v)$ for any $(z, z', x_v) \in \spaceX^2 \times \spaceX_v$ with $z_{\neigh(v)} = z_{\neigh(v)}'$, where $z_\myBlock \coloneqq (z_v)_{v \in \myBlock}$, for any $\myBlock \subseteq \vertexSet$. Under the model, the $v$th component thus only depends on the components in $\neigh(v)$ at the previous time step which allows us to slightly abuse notation to write $\smash{\trans_v(z_{\neigh(v)}, x_v) \coloneqq \trans_v(z, x_v)}$.
\end{myenumerate}
The structure of the model is illustrated in Fig.~\ref{fig:ssm}.

For the \glsdescplural{BPF} reviewed in this section, we also assume that the proposal kernel used by the \gls{PF} factorises in the same way as the model transitions. That is, $\smash{\prop_t(z, x) \coloneqq \prod_{v \in \vertexSet} \prop_{t,v}(z, x_v)}$, where $\prop_{t,v}\colon \spaceX \times \spaceX_v \to (0,\infty)$ are transition densities \WRT $\refMeasX_v$. We assume that $\prop_t$ induces the same spatial correlation structure as $\trans$ which allows us to write $\smash{\prop_{t,v}(z_{\neigh(v)}, x_v) \coloneqq \prop_{t,v}(z, x_v)}$. Again, all above-mentioned densities may depend on some model parameter $\theta$  though we suppress $\theta$ from the notation if possible.

\subsection{Localisation}

Let $\Filt_{t,\myBlock}$ denote the marginal of $\Filt_t$ on $\smash{\spaceX_\myBlock \coloneqq \prod_{v \in \myBlock} \spaceX_v}$. Assume that $\testFun \colon \spaceX \to \reals$ is \emph{local,} \IE it only depends on the components in some block $\myBlock \subseteq \vertexSet$. More formally, there exists $\smash{\testFun_\myBlock\colon \spaceX_\myBlock \to \reals}$ such that $\smash{\testFun(z) = \testFun(x) \eqqcolon \testFun_\myBlock(x_\myBlock)}$ for any $x,z \in \spaceX$ with $x_\myBlock = z_\myBlock$. In this case, $\smash{\Filt_t(\testFun) = \Filt_{t,\myBlock}(\testFun_\myBlock)}$.

The number of particles $\nParticles$ must usually grow exponentially in the model dimension $\modelDim$ to control the error of the \gls{PF} approximation in \eqref{eq:standard_pf_filter_approximation}. Unfortunately, this curse of dimensionality persists even if we only want to approximate the integral of a local function, $\smash{\Filt_t(\testFun) = \Filt_{t,\myBlock}(\testFun_\myBlock)}$, because the importance weights $W_t^n$ still depend on $\modelDim$.

A na\"ive attempt to stabilise the local filtering error, \IE the error of the particle approximation of $\Filt_{t,\myBlock}(\testFun_\myBlock)$, as $\modelDim$ grows, would be to replace the weights in \eqref{eq:standard_pf_filter_approximation} by weights which only depend on the components of the particles $X_t^{1:N}$ in $\myBlock$. That is, we could approximate $\Filt_{t,\myBlock}(\testFun_\myBlock)$ by
\begin{equation}
 \sum_{n = 1}^\nParticles W_{t,\myBlock}^n \testFun_\myBlock(X_{t,\myBlock}^n),\label{eq:standard_pf_marginal_filter_approximation}
\end{equation} 
with \emph{local} weights $\smash{W_{t,\myBlock}^n \coloneqq w_{t,\myBlock}^n / \sum_{m=1}^\nParticles w_{t,\myBlock}^m}$ defined through
\begin{align}
 w_{t,\myBlock}^n
 & = \smashoperator{\prod_{v \in \myBlock}} G_{t,v}(X_{t-1,\neigh(v)}^{A_{t-1}^n}, X_{t,v}^n) 
 = \smashoperator{\prod_{v \in \myBlock}} w_{t,v}^n.
\end{align}
Here, we have set $\neigh(\myBlock) \coloneqq \bigcup_{v \in \myBlock}\neigh(v)$, for any $\myBlock \subseteq \vertexSet$, and
\begin{align}                                                                                                                                                                                                                                                                                                                                                                         
  G_{t,v}(z_{\neigh(v)}, x_{v}) = \frac{\trans_{v}(z_{\neigh(v)}, x_{v})\obs_{v}(x_{v}, y_{t,v})}{\prop_{t,v}(z_{\neigh(v)}, x_{v})}.
\end{align}
The approximation in \eqref{eq:standard_pf_marginal_filter_approximation} is clearly independent of the dimension of $X_t$. Unfortunately, for $t>1$, it still suffers from the curse of dimensionality because of the particles and weights indirectly depend on (the dimension of) $X_{1:t-1}$.

\subsection{Blocked Particle Filters}
\label{subsec:bpf}
\glsreset{BPF}

Let $\myBlockSet$ be a partition of $\vertexSet$, \IE $\bigcup_{\myBlock \in \myBlockSet} = \vertexSet$ and $K \cap K' = \emptyset$ for all $K, K' \in \myBlockSet$. To stabilise the local filtering errors, the \emph{\gls{BPF}} from \citet{rebeschini2015local} seeks to make the entire evolution of the particle system  depend only on local weights $\smash{W_{t,\myBlock}^n = w_{t,\myBlock}^n / \sum_{i=1}^\nParticles w_{t,\myBlock}^i}$, for $\myBlock \in \myBlockSet$ -- and not just the construction of marginal filter approximations as in \eqref{eq:standard_pf_marginal_filter_approximation}.

Algorithm~\ref{alg:blocked_pf} summarises the \gls{BPF}. As before, we use the convention that any action described for \emph{some} $v$, $\myBlock$, or $n$ is to be performed conditionally independently for \emph{all} $v \in \vertexSet$, $\myBlock \in \myBlockSet$ and $n \in \{1,\dotsc, \nParticles\}$.

\begin{flushleft}
\begin{framedAlgorithm}[blocked particle filter \citep{rebeschini2015local}] \label{alg:blocked_pf}~
\begin{flushleft}
\begin{myenumerateAlg}
 \item At Step~$1$,
  \begin{myenumerateAlg}
    \item sample $\smash{X_{1}^n \sim \Prop_{1}}$,
    \item set $\smash{w_{1,v}^n \coloneqq G_{1,v}(X_{1,v}^n)}$ and $w_{1,\myBlock}^n \coloneqq \prod_{v\in\myBlock}w_{1,v}^n$.
  \end{myenumerateAlg}
 \item At Step~$t$, $t > 1$,
  \begin{myenumerateAlg}
    \item sample $\smash{A_{t-1, \myBlock}^n \sim \dCat(W_{t-1,\myBlock}^{1:\nParticles}})$,
    \item concatenate $\smash{\widebar{X}_{t-1}^n \coloneqq (X_{t-1,\myBlock}^{A_{t-1,\myBlock}^n})_{\myBlock \in \myBlockSet}}$,
    \item \label{alg:blocked_pf:2ii} sample $\smash{X_{t}^n \sim \Prop_{t}(\widebar{X}_{t-1}^n, \ccdot)}$,
    \item \label{alg:blocked_pf:2iii} set $\smash{w_{t,v}^n \coloneqq G_{t,v}(\widebar{X}_{t-1, \neigh(v)}^n, X_{t,v}^n)}$; $w_{t,\myBlock}^n \coloneqq \prod_{v\in\myBlock} w_{t,v}^n$.
   \end{myenumerateAlg}
 \end{myenumerateAlg}
\end{flushleft}
\end{framedAlgorithm}
\end{flushleft}

To obtain an approximation of the filter, \citet{rebeschini2015local}  construct the blocking approximation
\begin{align}
 \FiltTilde_{t,\mathsc{\bpf}}^\nParticles(\testFun) 
  \coloneqq \Bigl[\bigotimes_{\myBlock \in \myBlockSet} \FiltTilde^\nParticles_{t, \myBlock}\Bigr](\testFun), \label{eq:bpf_blocked_filt_approx}
\end{align}
where, for any $\myBlock \subseteq \vertexSet$ (\IE even for $\myBlock \notin \myBlockSet$), we have defined
\begin{equation}
\FiltTilde^\nParticles_{t,\myBlock}(\testFun_{\myBlock}) 
  \coloneqq \sum_{n = 1}^\nParticles W_{t,{\myBlock}}^n \testFun_\myBlock(X_{t,\myBlock}^n) . \label{eq:blocked_pf_marginal_filter_approximation}
\end{equation}
We use the `tilde'-symbol to stress that in contrast to standard \glspl{PF}, the \gls{BPF} asymptotically (as $\nParticles \to \infty$) does not target the true filter $\Filt_t$ but rather an approximate, blocked filter $\FiltTilde_t$ which is defined in Appendix~\ref{subsec:model_tilde}.

The complexity of the \gls{BPF} is $(\nParticles \maxBlockSize \card \myBlockSet)$ per time step where $\maxBlockSize \coloneqq \max_{\myBlock \in \myBlockSet} \card \myBlock$. Under strong mixing assumptions, \citet{rebeschini2015local, rebeschini2014comparison} show that local errors of $\smash{\FiltTilde_{t,\mathsc{\bpf}}^\nParticles}$ are bounded uniformly in $t$ and $\modelDim$.

\begin{remark}\label{rem:practical_bpf_approximation_full} Recall that the particle smoothers from Section~\ref{sec:standard_particle_filtering_and_smoothing} require approximations of the filters. Unfortunately,  approximating $\Filt_t$ by $\smash{\FiltTilde_{t,\mathsc{\bpf}}^\nParticles}$ has complexity $\smash{\bo(\nParticles^{\card \myBlockSet})}$ which is typically prohibitive. Instead, we propose to approximate $\Filt_t$ by subsampling $\nParticles$ points conditionally independently from $\smash{\FiltTilde_{t,\mathsc{\bpf}}^\nParticles}$. This has complexity $\bo(\nParticles)$.
\end{remark}

\begin{remark}\label{rem:practical_bpf_approximation_marginal}
 The blocked particle smoothers proposed in Section~\ref{sec:methodology}, will require approximations of filter marginals on blocks $\smash{\myBlock' \supseteq \myBlock \in \myBlockSet}$. More specifically, $\myBlock'$ will be some neighbourhood of $\myBlock$. Unfortunately, approximating $\smash{\Filt_{t, \myBlock'}}$ by marginalising $\smash{\FiltTilde_{t,\mathsc{\bpf}}^\nParticles}$ has complexity $\smash{\bo(\nParticles^{\card \mathcal{L}})}$, where $\smash{\mathcal{L} \coloneqq \{\myBlock \in \myBlockSet \,|\, \myBlock \cap \myBlock' \neq \emptyset\}}$, and this is typically prohibitive. Instead, we propose to approximate $\smash{\Filt_{t, \myBlock'}}$ by $\smash{\FiltTilde^\nParticles_{t,\myBlock'}}$ (\IE via \eqref{eq:blocked_pf_marginal_filter_approximation} with $\myBlock' = \myBlock$). This has complexity $\bo(\nParticles)$.
\end{remark}

\section{Blocked Particle Smoothing}
\label{sec:methodology}

In this section, we formally show that standard particle smoothing breaks down in high dimensions -- even under dimensionally stable filter approximations. We then propose novel blocked particle smoothers which provably circumvent this curse of dimensionality in the canonical high-dimensional model from Subsection~\ref{subsec:high-dimensional_ssm} (potential extensions to other models are discussed in Subsection~G of the supplementary materials).

\subsection{High-dimensional Smoothing}
\label{subsec:high-dimensional_smoothing}
For the remainder of this work, we will be concerned with approximating smoothed expectations $\smooth_t(\smoothFun_t)$ for the canonical high-dimensional model in the case that $\smoothFun_t\colon \spaceX^t \to \reals$ is additive in both \emph{time and space}\footnote{Strictly speaking, as with standard particle smoothers, additivity in \emph{time} in only needed for the \gls{FS} but not the \gls{BS} variant of the algorithm.}, in the sense that there exist functions $\testFun_{1,v} \colon \spaceX_v \to \reals$ and $\testFun_{t,v} \colon \spaceX_{\neigh(v)} \times \spaceX_v \to \reals$, for $t > 1$, such that 
\begin{align}
  \smoothFun_t(x_{1:t}) 
  = \sum_{s = 1}^t \sum_{v \in \vertexSet} \testFun_{s,v}(x_{s-1, \neigh(v)},x_{s,v}), \label{eq:additive_in_time_and_space}
\end{align}
with the convention that any quantity with time index $0$ is to be ignored. Each constituent function $\testFun_{t,v}$ may also implicitly depend on $y_{t,v}$. 

We now give an example of such an additive function which is important for the problem of calibrating $\theta$.

\begin{example}[score]\label{ex:score}
 Assume that $\init_v = \init_v^\theta$, $\trans_v=\trans_v^\theta$, $\obs_v=\obs_v^\theta$ and hence $\uSmooth_\nTimeSteps = \uSmooth_\nTimeSteps^\theta$ depend on some unknown parameter vector $\theta$. Approximating the (marginal) \gls{MLE} of $\theta$ via a gradient-ascent algorithm requires computing the score, \IE the gradient of the (marginal) log-likelihood, given by $\smash{\nabla_\vartheta \log \uSmooth_\nTimeSteps^\vartheta(\unitFun)|_{\vartheta = \theta} = \smooth_\nTimeSteps^\theta(\smoothFun_\nTimeSteps^\theta)}$, where $\smoothFun_\nTimeSteps^\theta$ is as in \eqref{eq:additive_in_time_and_space} with 
\begin{gather}
 \testFun_{1,v}^\theta(x_{v})
 = \nabla_\vartheta \log \bigl[\init_v^\vartheta(x_{v}) \obs_v^\vartheta(x_{v}, y_{1,v})\bigr]\bigr|_{\vartheta = \theta}, \label{eq:additive_functional_gradient_1}\\*
 \testFun_{t,v}^\theta(z_{\neigh(v)}, x_{v}) = \nabla_\vartheta \log \bigl[\trans_v^\vartheta(z_{\neigh(v)}, x_{v}) \obs_v^\vartheta(x_{v}, y_{t,v})\bigr]\bigr|_{\vartheta = \theta}. \label{eq:additive_functional_gradient_2}
\end{gather}
\end{example}


\subsection{Assumptions}
\label{subsec:assumptions}

In this subsection, we state some assumptions under which we prove various theoretical results below. 

Assumption~\ref{as:mixing} is a regularity condition routinely used in the analysis of \gls{SMC} techniques (\EG\citep{delmoral2004feynman}). It can often be relaxed at the price of significantly complicating the analysis \citep{douc2014long, whiteley2013stability}. Assumption~\ref{as:local_test_function} requires $\smoothFun_\nTimeSteps$ to be spatially and temporally local. Recall that by \eqref{eq:additive_in_time_and_space}, the smoothing functional of interest decomposes into a sum of such local functions. For simplicity, as in \citet{delmoral2010forward, delmoral2010backward}, we assume here that $\testFun_{t,J}$ does not depend on the state at time~$t-1$ but this could be relaxed. Assumption \ref{as:filter_approximation} lists a number of options for the Monte Carlo approximation $\Filt_t^\nParticles$ of the filter $\Filt_t$ needed for the standard particle smoothers. 


\begin{assumption}[strong mixing condition]~\label{as:mixing}
 For any $v \in \vertexSet$ the dominating measure $\refMeasX_v$ is finite, and there exist $\varepsilon > 0$ such that for all $(x,z,y) \in \spaceX^2 \times \spaceY$ and all $v \in \vertexSet$,
 \begin{align}
  \varepsilon^{-1}\leq \trans_v(x_{\neigh(v)}, z_v) \leq \varepsilon \;\; \text{and} \;\; 0 < \!\int_{\spaceX_v} \!\!\!\obs_v(z_v, y_v) \psi_v(\diff z_v) < \infty.\!\!\!\!
 \end{align}
\end{assumption}

\begin{assumption}[local test function]\label{as:local_test_function}
  There exist $J \subseteq \myBlock \in \myBlockSet$, $r \in \{1,\dotsc,\nTimeSteps\}$, and $\testFun_{r,J}\colon \spaceX_{J} \to \reals$ with $\lVert \testFun_{r,J} \rVert  \leq 1$ such that $\testFun_{r,J}$ is not $\refMeasX$-almost everywhere constant and
  \begin{equation}
    \smoothFun_{\nTimeSteps}(x_{1:\nTimeSteps}) = \testFun_{r,J}(x_{r,J}), \quad \text{for all $x_{1:\nTimeSteps} \in \spaceX^{\nTimeSteps}$.}
  \end{equation}
\end{assumption}


\glsunset{IID}

\begin{assumption}[filter approximations]~\label{as:filter_approximation} For any $t \in \naturals$, approximate $\Filt_t$ using
  \begin{myenumerate}[label=\alph*), ref=\alph*]
    \item \label{as:filter_approximation:pf} a standard \gls{PF} with $\nParticles$ particles  (\IE via \eqref{eq:standard_pf_filter_approximation}),
    
    
    \item \label{as:filter_approximation:bpf} $\nParticles$ samples drawn conditionally independently from the \gls{BPF} approximation in \eqref{eq:bpf_blocked_filt_approx} (see Remark~\ref{rem:practical_bpf_approximation_full}),
    

%
%
     \item \label{as:filter_approximation:pfIid} $\nParticles$ \gls{IID} samples from the exact filter, $\Filt_{t}$,
  
     \item \label{as:filter_approximation:bpfIid} $\nParticles$ \gls{IID} samples from the blocked filter $\FiltTilde_{t}$ (see Appendix~\ref{subsec:model_tilde}).
    
  \end{myenumerate}
\end{assumption}

\subsection{Breakdown of Standard Particle Smoothing}
\label{subsec:motivation}

In this subsection, we show that standard particle smoothing suffers from a curse of dimensionality, even if local filter errors are dimension-independent. 

The efficiency of standard particle smoothing relies strongly on the mixing properties of the transitions $\trans$. Unfortunately, the mixing of $\trans(x,z) = \prod_{v \in \vertexSet}\trans_v(x,z_v)$ degrades exponentially in the model dimension $\modelDim$, unless the local transitions are perfectly mixing in the following sense.

\begin{definition}[perfect mixing] \label{def:perfect_mixing}
 The local transitions $\trans_v$ are called \emph{perfectly mixing} if $\int_A \trans_v(x_{\neigh(v)}, z_v) \refMeasX_v(\diff z_v)$ is $\prod_{u\in \neigh(v)}\refMeasX_u$-almost everywhere constant for all $A \subseteq \spaceX_v$. 
\end{definition}

Note that if the transitions are perfectly mixing, the latent states are independent over time in which case particle filtering/smoothing methodology is not needed.

Our main result in this subsection is Proposition~\ref{prop:exploding_asymptotic_variance}, proved in Appendix~\ref{app:proof_of_exploding_variance}. It shows that the asymptotic variance associated with the standard \gls{FS} approximation $\smooth^\nParticles_{\nTimeSteps,\fs}(\smoothFun_\nTimeSteps)$, defined in \eqref{eq:standard_forward_smoothing_approximation} (see also Algorithm~\ref{alg:standard_fs}), grows exponentially in the model dimension unless the model transitions are perfectly mixing. The result holds even though we assume highly favourable conditions, summarised in Assumption~\ref{as:spatially_iid_model}, and even if local filter errors are dimension-independent.

\begin{assumption}[spatially IID model]\label{as:spatially_iid_model} We have $\neighSize = 0$, \IE $\smash{\trans(x, z) = \prod_{v \in \vertexSet} \trans_v(x_v, z_v)}$, for any $(x,z) \in \spaceX^2$. In addition, for any $t \leq \nTimeSteps$ and any $u,v \in \vertexSet$, we have $\spaceX_u = \spaceX_v \eqqcolon \spaceXComp$, $\spaceY_u = \spaceY_v \eqqcolon \spaceYComp$, $\trans_u(x, z) = \trans_v(x, z) \eqqcolon \transComp(x,z)$, $\obs_u(x, y_{t,u}) = \obs_v(x, y_{t,v}) \eqqcolon \obsComp_t(x)$, $\refMeasX_u = \refMeasX_v \eqqcolon \refMeasXComp$ and $\refMeasY_u = \refMeasY_v \eqqcolon \refMeasYComp$. 
\end{assumption}

We stress that we only use Assumption~\ref{as:spatially_iid_model} (which implies a complete absence of spatial interactions) to highlight the fact that standard particle smoothers break down in high dimensions even under such highly favourable conditions. The novel smoothing methodology proposed below does not rely on this assumption.

\begin{proposition}\label{prop:exploding_asymptotic_variance}~
 \begin{myenumerate}
  \item \label{prop:exploding_asymptotic_variance:1} Under Assumptions~\ref{as:mixing}, \ref{as:local_test_function}, \ref{as:spatially_iid_model} and if the filter is approximated according to Assumption~\ref{as:filter_approximation}\ref{as:filter_approximation:pf} (using a bootstrap \gls{PF}), or according to Assumption~\ref{as:filter_approximation}\ref{as:filter_approximation:pfIid},
 \begin{equation}
  \sqrt{N}\bigl[\smooth^\nParticles_{\nTimeSteps,\fs}(\smoothFun_\nTimeSteps) - \smooth_\nTimeSteps(\smoothFun_\nTimeSteps)\bigr]\Rightarrow \dN(0, \sigma_\nTimeSteps^2(\smoothFun_\nTimeSteps)), 
\end{equation}
 as $\nParticles \to \infty$, where 
 \begin{equation}
  \sigma_{\nTimeSteps}^2(\smoothFun_\nTimeSteps) \geq \sum_{t=1}^{\smash{T}} a_{t, \nTimeSteps}^{\vphantom{V}}(\testFun_{r,J}) (c_{t, \nTimeSteps})^\modelDim. \label{eq:asymptotic_variance_in_proposition}
 \end{equation}
 Here, for each $t \leq \nTimeSteps$, $a_{t,\nTimeSteps}(\testFun_{r,J}) > 0$ and $c_{t,\nTimeSteps} \geq 1$ do not depend on the model dimension, $\modelDim$.
 
 \item \label{prop:exploding_asymptotic_variance:2} For all $t \leq \nTimeSteps$, $c_{t, \nTimeSteps} > 1$, unless the model transitions are perfectly mixing.
 \end{myenumerate}
\end{proposition}

Note that $c_{t, \nTimeSteps} > 1$ implies that the asymptotic variance in Equation~\refeq{eq:asymptotic_variance_in_proposition} grows exponentially in $\modelDim$. 

Although Proposition~\ref{prop:exploding_asymptotic_variance} has been established for \gls{FS} (which computes the sum in \eqref{eq:standard_forward_smoothing_approximation} exactly), it extends immediately to \gls{BS} and to the \gls{PaRIS} algorithm from \citet{olssonEfficient} as these construct sampling approximations of \eqref{eq:standard_forward_smoothing_approximation}. Due to \eqref{eq:fs_as_rao-blackwellised_bs}, they cannot attain a smaller variance than \gls{FS} (for equal numbers of particles $\nParticles$).

%

\subsection{Proposed Algorithms}
\label{subsec:proposed_algorithm}

In this subsection, we propose novel `blocked' particle smoothers aimed at circumventing the curse of dimensionality analysed in Proposition~\ref{prop:exploding_asymptotic_variance}.

\subsubsection{Blocked Backward Kernels} 
The curse of dimensionality suffered by standard particle smoothing methods is due to the dependence of the backward kernel in \eqref{eq:backward_kernel} on the mixing properties of the full model transitions $\trans$ (as outlined above, these deteriorate as the model dimension increases). Thus, we need to design approximate versions of the backward kernels which only operate on some `fixed' block $\myBlock \subseteq \vertexSet$ of components and which therefore only rely on the mixing properties of the transitions associated with $\myBlock$:
\begin{equation}
 \trans_{\myBlock}(x_{\neigh(\myBlock)},z_{\myBlock}) \coloneqq \prod_{v \in {\myBlock}}\trans_v(x_{\neigh(v)},z_v).
\end{equation}
To achieve this, we employ \emph{blocked} backward kernels $\back_{{\myBlock}, \nu}\colon \spaceX_{\myBlock} \times \Borel(\spaceX_{\neigh(\myBlock)}) \to [0,1]$, for $x \in \spaceX_{\myBlock}$ given by
\begin{equation}
 \back_{{\myBlock}, {\Filt_{t,\neigh({\myBlock})}}}(x, \diff z)
 \coloneqq 
 \frac{\trans_{{\myBlock}}(z, x) {\Filt_{t,\neigh({\myBlock})}}( \diff z)}{\int_{\spaceX_{\neigh(\myBlock)}} \trans_{{\myBlock}}(u, x){\Filt_{t,\neigh({\myBlock})}}(\diff u)}. \label{eq:blocked_backward_kernel}
\end{equation}
Again, the filter marginal $\smash{\Filt_{t,\neigh({\myBlock})}}$ is typically intractable and must be replaced by a Monte Carlo approximation $\Filt_{t,\neigh({\myBlock})}^N$.


\subsubsection{Blocked Forward Smoothing}
In the remainder of this work, for any $\myBlock \in \myBlockSet$, we let $\myBlock \subseteq \enlargedBlock \subseteq \vertexSet$ be some enlarged block containing $\myBlock$. Note that with this notation, $\smash{\Filt^\nParticles_{t,\neigh(\enlargedBlock)}}$ is an approximation of the marginal filter on the components in the neighbourhood of the enlargement of $\myBlock$.


Our proposed blocked \gls{FS} scheme is outlined in Algorithm~\ref{alg:blocked_fs}, where as usual, any action prescribed for \emph{some} $n$ is to be performed conditionally independently for \emph{all} $n \in \{1, \dotsc, \nParticles\}$.

\begin{flushleft}
\begin{framedAlgorithm}[blocked forward smoothing] 
\begin{flushleft}
\label{alg:blocked_fs}~
 \begin{myenumerateAlg}
  \item Perform the following steps (in parallel) for any $\myBlock \in \myBlockSet$.
  \begin{myenumerateAlg}
  \item Set $\alpha_{1,\enlargedBlock}^n \coloneqq \testFun_{1,\myBlock}(X_{1,\myBlock}^n)$. For $t > 1$, set
  \begin{align}
   \!\!\!\!\!\!\!\!\!\!\!\!\!\!\alpha_{t+1,\enlargedBlock}^n 
   & \coloneqq \sum_{\smash{m = 1}}^{\smash{\nParticles}} \back_{\enlargedBlock, \Filt^\nParticles_{t,\neigh(\enlargedBlock)}}(X_{t+1,\enlargedBlock}^n, \{X_{t,\neigh(\enlargedBlock)}^m\})
   \\[-1ex]
   & \qquad \qquad\times \bigl[\alpha_{t,\enlargedBlock}^m + \testFun_{t+1,\myBlock}(X_{t,\neigh(\myBlock)}^m, X_{t+1,\myBlock}^n)\bigr].
  \end{align}
 \end{myenumerateAlg}
  \item Approximate $\smoothFunIntegral_t$ by $\smash{\smoothFunIntegral^\nParticles_t \coloneqq \sum_{\myBlock \in \myBlockSet} \smoothFunIntegral^\nParticles_{t, \myBlock}}$, where, for any $\myBlock \in \myBlockSet$,  $\smash{\smoothFunIntegral^\nParticles_{t, \myBlock} \coloneqq \sum_{n=1}^\nParticles W_{t,\enlargedBlock}^n \alpha_{t,\enlargedBlock}^n}$.
 \end{myenumerateAlg}
\end{flushleft}
\end{framedAlgorithm}
\end{flushleft}

As in the case of standard \gls{FS} (Algorithm~\ref{alg:standard_fs}), Algorithm~\ref{alg:blocked_fs} can be implemented online. That is, the terms $\alpha_{t,\enlargedBlock}^n$ can be determined at the $t$th step of the Monte Carlo filter used to generate $(X_{t,v}^n, w_{t,v}^n)$. The computational complexity is $\bo(\nParticles^2 \nTimeSteps [\myBlockSetSize] \max_{\myBlock \in \myBlockSet} \card \neigh(\enlargedBlock) )$ when running the algorithm up to time~$T$. This is slightly higher than that of the (dimensionally unstable) Algorithm~\ref{alg:standard_fs}. However, significant speed-ups should be attainable since the blocked smoothing recursions can be run in parallel on distributed architectures or multiple cores (see~\citep{lee2010utility}).

\subsubsection{Blocked Backward Sampling}
As with standard particle smoothing, we may reduce the computational complexity of blocked \gls{FS} to $\bo(\nParticles \nBackwardPaths \nTimeSteps [\myBlockSetSize] \max_{\myBlock \in \myBlockSet} \card \neigh(\enlargedBlock) )$ via a simple Monte Carlo approximation based on $\nBackwardPaths < \nParticles$ particle paths. Algorithm~\ref{alg:blocked_bs} outlines blocked \gls{BS}. Here, we use the convention that any action prescribed for \emph{some} $m$ is to be performed conditionally independently for \emph{all} $m \in \{1, \dotsc, \nBackwardPaths\}$. 

\begin{flushleft}
\begin{framedAlgorithm}[blocked backward sampling]
\begin{flushleft}
\label{alg:blocked_bs}~
\begin{myenumerateAlg}
  \item Perform the following steps (in parallel) for any $\myBlock \in \myBlockSet$.
 \begin{myenumerateAlg}
 \item Sample $\smash{\widetilde{X}_{\nTimeSteps,\enlargedBlock}^m \sim \Filt^\nParticles_{\nTimeSteps,\enlargedBlock}}$. For $t = \nTimeSteps-1,\dotsc,1$, sample
 \begin{equation}
  \smash{\widetilde{X}_{t,\neigh(\enlargedBlock)}^m \sim \back_{\enlargedBlock, \Filt^\nParticles_{t,\neigh(\enlargedBlock)}}(\widetilde{X}_{t+1, \enlargedBlock}^m, \ccdot)}.
 \end{equation}

 \item \label{alg:blocked_bs:1:iv} Set $\smoothFunIntegral^\nParticles_{\nTimeSteps, \myBlock}
  \coloneqq \frac{1}{\nBackwardPaths}\sum_{m = 1}^\nBackwardPaths \sum_{t=1}^\nTimeSteps \testFun_{t,\myBlock}(\widetilde{X}_{t-1,\neigh(\myBlock)}^m, \widetilde{X}_{t,\myBlock}^m)$.
 \end{myenumerateAlg}
  \item Approximate $\smoothFunIntegral_\nTimeSteps$ by $\smoothFunIntegral^\nParticles_{\nTimeSteps} \coloneqq \sum_{\myBlock \in \myBlockSet} \smoothFunIntegral^\nParticles_{\nTimeSteps, \myBlock}$.
 \end{myenumerateAlg}
\end{flushleft}
\end{framedAlgorithm}
\end{flushleft}

\subsubsection{Bias Reduction} The blocking strategy introduces a bias but as shown in Proposition~\ref{prop:asymptotic_bias} below, this bias is \emph{bounded} uniformly in the time horizon and model dimension. In addition, we can \emph{reduce} the bias by defining the enlarged blocks as $\smash{\enlargedBlock \coloneqq \neigh_i(\myBlock) \coloneqq \bigcup_{v \in \myBlock} \neigh_i(v)}$, for some $i>0$. That way, blocked particle smoothers do not require evaluating test functions at components near block boundaries (by Proposition~\ref{prop:asymptotic_bias} below, the bias decays exponentially in the distance to the block boundary). 

Though as we will discuss in more detail in Subsection~\ref{subsec:theory}, any bias reduction attained by choosing larger (enlarged) blocks needs to be carefully balanced against the variance increase this induces. In particular, taking $\enlargedBlock = \vertexSet$ trivially minimises the bias but then the blocked particle smoothers coincide with the (dimensionally unstable) standard particle smoothers.

\subsubsection{Marginal Filter Approximations}
As with standard particle smoothers, any stable (Monte Carlo) approximation of filter marginals can be plugged into Algorithms~\ref{alg:blocked_fs} and \ref{alg:blocked_bs}. We consider the following approximations $\Filt_{t,\myBlock'}^\nParticles$ of filter marginals $\Filt_{t,\myBlock'}$.
%
%
\glsunset{IID}

\begin{assumption}[marginal filter approximations]~\label{as:marginal_filter_approximation}  \!\!\!\!For any $t \in \naturals$ and any block $\myBlock' \subseteq \vertexSet$, approximate $\Filt_{t,\myBlock'}$ using a
\begin{myenumerate}[label=\alph*), ref=\alph*]

 \item \label{as:marginal_filter_approximation:pf} standard \gls{PF} but based on local weights as in \eqref{eq:standard_pf_marginal_filter_approximation},

 \item \label{as:marginal_filter_approximation:bpf} \gls{BPF} (via  \eqref{eq:blocked_pf_marginal_filter_approximation} as justified in Remark~\ref{rem:practical_bpf_approximation_marginal}).
 
 \item \label{as:marginal_filter_approximation:pfIid} suitable marginal of Assumption~\ref{as:filter_approximation}\ref{as:filter_approximation:pfIid} (\gls{IID} samples from $\Filt_t$),
  
 \item \label{as:marginal_filter_approximation:bpfIid} suitable marginal of Assumption~\ref{as:filter_approximation}\ref{as:filter_approximation:bpfIid} (\gls{IID} samples from $\FiltTilde_t$).
\end{myenumerate}
\end{assumption}
 
%
%

\subsection{Theoretical Analysis}
\label{subsec:theory}

%
%

In this subsection, we do not consider enlarged blocks, \IE the blocks used for smoothing are the same blocks employed by the \gls{BPF}. Under Assumption~\ref{as:local_test_function}, for $J \subseteq \myBlock = \enlargedBlock \in \myBlockSet$, the blocked \gls{FS} approximation of $\smooth_\nTimeSteps(\smoothFun_\nTimeSteps)$ then simplifies to
\begin{align}
 \smooth^\nParticles_{\nTimeSteps,\fsBlock{\myBlock}}(\smoothFun_\nTimeSteps) 
  & = \int \testFun_{r,J}(x_{r,J}) \Filt_{\nTimeSteps, {\myBlock}}^\nParticles(\diff x_{\nTimeSteps, {\myBlock}})\\* 
  & \quad \times \smashoperator{\prod_{t=1}^{\smash{\nTimeSteps-1}}} \back_{{\myBlock}, \Filt_{t,\neigh({\myBlock})}^\nParticles}(x_{t+1,{\myBlock}}, \diff x_{t,\neigh({\myBlock})}). \label{eq:blocked_fs_approx}
\end{align}
We now derive uniform (in both the time horizon $\nTimeSteps$ and model dimension $\modelDim$) bounds on the asymptotic (as $\nParticles \to \infty$) bias and variance of the blocked \gls{FS} estimator $\smooth^\nParticles_{\nTimeSteps,\fsBlock{\myBlock}}(\smoothFun_\nTimeSteps)$. Following \citet{rebeschini2015local}, we also show that this estimator can be made locally consistent by scaling $\myBlock$ appropriately with $\nParticles$. 

\subsubsection{Variance} 
We now state Proposition~\ref{prop:asymptotic_variance}, proved in Appendix~\ref{app:proofs_of_variance_and_bias_bounds}, which suggests that the variance of blocked \gls{FS} is bounded in time and in the model dimension but may grow exponentially in the block size for a fixed number of particles (so that controlling the variance requires $N$ to grow exponentially in the block size). To prove this result, we specifically assume that we approximate the filter at any time~$t$ using \gls{IID} samples from $\FiltTilde_t$ (Part~\ref{prop:asymptotic_variance:1}) or from $\Filt_t$ (Part~\ref{prop:asymptotic_variance:2}). In addition, recall that $\maxBlockSize = \max_{\myBlock \in \myBlockSet} \card \myBlock$ and let $\dN(\mu, \sigma^2)$ be a normal distribution with mean $\mu \in \reals$ and variance $\sigma^2>0$. The measures $\smoothBar_\nTimeSteps$ and $\smoothHat_\nTimeSteps$ which govern the asymptotic mean in this central limit theorem are specified in Appendix~\ref{app:proofs_of_variance_and_bias_bounds}. We stress that Part~\ref{prop:asymptotic_variance:2} of this proposition is included mainly as a direct contrast to the negative result in Proposition~\ref{prop:exploding_asymptotic_variance}. 

\begin{proposition}[asymptotic variance]\label{prop:asymptotic_variance}
 Under Assumptions~\ref{as:mixing} and \ref{as:local_test_function}, there exist probability measures $\smoothBar_\nTimeSteps$ and $\smoothHat_\nTimeSteps$ on $\spaceX^\nTimeSteps$ such that the following statements hold.
 \begin{myenumerate}
  \item\label{prop:asymptotic_variance:1} If filter marginals are approximated per Assumption~\ref{as:marginal_filter_approximation}\ref{as:marginal_filter_approximation:bpfIid}
  \begin{equation}
  \sqrt{\nParticles}\bigl[\smooth_{\nTimeSteps,\fsBlock{\myBlock}}^\nParticles(\smoothFun_\nTimeSteps) - \smoothBar_\nTimeSteps(\smoothFun_\nTimeSteps)\bigr] \Rightarrow \dN(0, \sigma_\nTimeSteps^2(\smoothFun_\nTimeSteps)), 
  \end{equation}
  as $\nParticles \to \infty$, where for $\consta > 0$ which only depends on $\varepsilon$,
  \begin{equation}
    \sigma_\nTimeSteps^2(\smoothFun_\nTimeSteps) \leq \eul^{\consta \card\myBlock} \leq \eul^{\consta \maxBlockSize}.
  \end{equation}
  \item\label{prop:asymptotic_variance:2} Part~\ref{prop:asymptotic_variance:1} remains valid (but with asymptotic mean $\smoothHat_\nTimeSteps(\smoothFun_\nTimeSteps)$) if filter marginals are approximated per Assumption~\ref{as:marginal_filter_approximation}\ref{as:marginal_filter_approximation:pfIid}. 
 \end{myenumerate}
\end{proposition}

\subsubsection{Bias}
We now state Proposition~\ref{prop:asymptotic_bias}, proved in Appendix~\ref{app:proofs_of_variance_and_bias_bounds}, which shows that the bias of blocked \gls{FS} is bounded in time and in the model dimension and decays exponentially in the distance to the block boundary. This result assumes that we use either the \gls{BPF} or \gls{IID} samples from $\FiltTilde_t$ to approximate the filter at each time step. In addition, for any $\myBlock, \myBlock' \subseteq \vertexSet$, we write $\metric(\myBlock, \myBlock') \coloneqq \min_{v \in \myBlock} \min_{v' \in \myBlock'} \metric(v,v')$ and $\partial \myBlock \coloneqq \{v \in \myBlock\mid \neigh(v) \nsubseteq \myBlock\}$. 

\begin{proposition}[asymptotic bias]\label{prop:asymptotic_bias}
 Under Assumptions~\ref{as:mixing}, \ref{as:local_test_function}, and if filter marginals are approximated either 
 per Assumption~\ref{as:marginal_filter_approximation}\ref{as:marginal_filter_approximation:bpf}
 or per Assumption~\ref{as:marginal_filter_approximation}\ref{as:marginal_filter_approximation:bpfIid},
 \begin{equation}
 \bigl\lvert \smoothBar_{\nTimeSteps}(\smoothFun_\nTimeSteps) - \smooth_\nTimeSteps(\smoothFun_\nTimeSteps)\bigr\rvert 
 \leq \constb \card(J) \eul^{- \constc \metric(J, \partial \myBlock)},
\end{equation}
 where $\constb \in \reals$ and $\constc > 0$ only depend on $\varepsilon$ and $\neighSize$.
\end{proposition}

\subsubsection{Consistency} 
Propositions~\ref{prop:asymptotic_variance} and \ref{prop:asymptotic_bias} indicate a bias--variance trade-off: for a fixed number of particles, the bias can be reduced by increasing the size of the blocks but the variance typically grows exponentially in the block size. To minimise the \gls{MSE}, we must therefore scale the size of the blocks suitably in $\nParticles$ to balance the variance and the (squared) bias.

For any $(t,v) \in \{1,\dotsc,\nTimeSteps\} \times \vertexSet$, define the function $\smash{\smoothFun_{t,v}\colon \spaceX^\nTimeSteps \to \reals}$ for any $x_{1:\nTimeSteps} \in \spaceX^\nTimeSteps$ by $\smash{\smoothFun_{t,v}(x_{1:\nTimeSteps}) \coloneqq x_{t,v}}$. The \emph{local} \gls{MSE} for component $v \in \myBlock \in \myBlockSet$ of blocked \gls{FS} is
\begin{equation}
 \MSE_{t,v}(\nParticles) \coloneqq \E\bigl[(\smooth_{\nTimeSteps,\fsBlock{\myBlock}}^\nParticles(\smoothFun_{t,v})- \smooth_{\nTimeSteps}(\smoothFun_{t,v}))^2\bigr].
\end{equation}
The blocked particle smoothing estimate can then be made locally \gls{MSE}-consistent by growing blocks suitably logarithmically in $\nParticles$ because Propositions~\ref{prop:asymptotic_variance} and \ref{prop:asymptotic_bias} imply
\begin{align}
 \!\!\MSE_{t,v}(\nParticles) = \bo\bigl( \eul^{- 2\constc \metric(v, \partial \myBlock)} + \nParticles^{-1}\eul^{\consta \card\myBlock}\bigr). \label{eq:local_mse}
\end{align}

As a simple illustration, assume that the spatial graph $\graphSet$ is a one-dimensional lattice as in Fig.~\ref{fig:ssm} and consider a component $v$ at the centre of $\myBlock \coloneqq \neigh_i(v)$ so that $\metric(v,\myBlock) = i$ and $\card\myBlock = 2i + 1$, for some $i \geq 0$. The local \gls{MSE} in \eqref{eq:local_mse} then vanishes as $N \to \infty$ if the block radius grows as $i= o([\log(N)/\consta - 1]/2)$. For instance, taking $i = \lfloor \log(N)/(8\consta)\rfloor$ implies that the local \gls{MSE} is $\bo(\exp(-\frac{\constc}{2\consta} \log(N)))$. A straightforward modification of \citet[Corollary~2.5]{rebeschini2015local} extends such local consistency to components $v$ which are not at the centre of some block or to the case that $\graphSet$ is a $q$-dimensional lattice. Though, in both cases, $i$ must grow even more slowly with $\nParticles$.

\section{Simulations}
\label{sec:simulations}

In this section, we compare standard and blocked particle smoothers on a high-dimensional state-space model. 

\subsection{The Model}

Blocking strategies have already been successfully applied to perform \emph{filtering} in a \gls{FMRI} application \citep{besada2009parallel} and in military multiple-target tracking scenarios \citep{das2005factored}. However, to assess the performance of our \emph{smoothing} algorithms, we consider the more abstract model from \citet{rebeschini2015local} which is increasingly popular as a benchmark for \gls{SMC} algorithms in high dimensions \citep{beskos2017stable, naesseth2015nested, guarniero2016iterated}. Purely in order to compare our method against analytical solutions, we let the model be linear and Gaussian. 


Let $\dN(\ccdot; \mu, \varSigma)$ denote the density of a normal distribution with suitable mean vector $\mu$ and covariance matrix $\varSigma$, let $\mathbf{0}_\modelDim \in \reals^\modelDim$ be a vector of zeros and let $\mathbf{I}_\modelDim \in \reals^{\modelDim \times \modelDim}$ be the identity matrix. Then the model is given by $\spaceX = \spaceY = \reals^\modelDim$, $\init^\theta(x) = \dN(x; \mathbf{0}_\modelDim, \mathbf{I}_\modelDim)$, 
\begin{gather}
 \trans^\theta(z, x) = \textstyle{\dN(x; Az, \sigma_X^2 \mathbf{I}_\modelDim),} \quad \obs^\theta(x, y) = \dN(y; x, \sigma_Y^2\mathbf{I}_\modelDim).
\end{gather}
%
Here, $\sigma_X, \sigma_Y > 0$ and $A = (a_{i,j})_{(i,j) \in \vertexSet^2}$ is a symmetric, banded diagonal (\IE symmetric Toeplitz) matrix whose diagonal entries are $a_0, a_1, \dotsc, a_\neighSize > 0$ for some $\neighSize \in \naturals \cup \{0\}$:
\begin{equation}
a_{i,j} \coloneqq
 \begin{cases}
  a_r,& \text{if $r \in \{0,1, \dotsc, \neighSize\}$ and $j \in \{i+r, i-r\}$,}\\
  0, & \text{otherwise.}
 \end{cases}
\end{equation}
This induces a local spatial correlation structure because 
\begin{align}
\trans_v^\theta(z_{\neigh(v)}, x_v)&= \textstyle{\dN(x_v; \sum_{r=0}^\neighSize a_r\sum_{u \in \calB_r(v)}z_u, \sigma_X^2),}
\end{align}
where $\calB_r(v) \coloneqq \{u \in \vertexSet\mid \metric(u,v) = r\}$ 
denotes the vertices in $\vertexSet$ whose distance from vertex~$v$ is exactly $r$. 

We parametrise the model via
\begin{equation}
 \theta \coloneqq \theta_{0:\neighSize+2} \coloneqq (a_0, a_1, \dotsc, a_\neighSize, \log \sigma_X, \log \sigma_Y).
\end{equation}
All simulation results use $\neighSize=1$ with true parameter values $a_0 = 0.5$, $a_1 = 0.2$, $\sigma_X=\sigma_Y=1$. 

\subsection{Smoothing}
\label{subsec:smoothing}

In this subsection, for fixed $\theta$, we estimate the smoothed sufficient statistic $\smoothFunIntegral_\nTimeSteps$, defined according to \eqref{eq:additive_in_time_and_space} with $\testFun_{1,v} \equiv 0$ and, for $t \in \naturals$, by
\begin{equation}
 \testFun_{t+1,v}(x_{t,\neigh(v)}, x_{t+1,v}) \coloneqq x_{t+1,v} \textstyle \sum_{u \in \calB_r(v)} x_{t,u}.
\end{equation}
A full list of sufficient statistics for this model is given in Subsection~I of the supplementary materials. 

\begin{figure}[!t]
  \noindent{}
  \centering
  \begin{subfigure}[t]{\linewidth}
    \centering
  \includegraphics[trim=0.5cm 0cm 0cm 0.4cm]{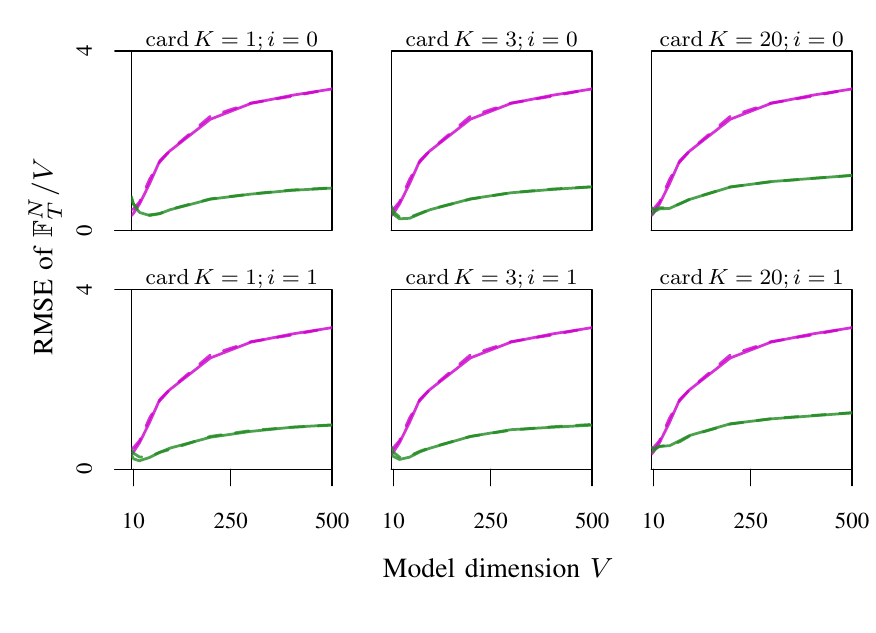}
    \caption{Using a standard \glsdesc{PF} to approximate the filter.}\label{fig:rmse_smc}		
  \end{subfigure}
  \\
  \quad
  \begin{subfigure}[t]{\linewidth}
    \centering
  \includegraphics[trim=0.5cm 0cm 0cm 0.4cm]{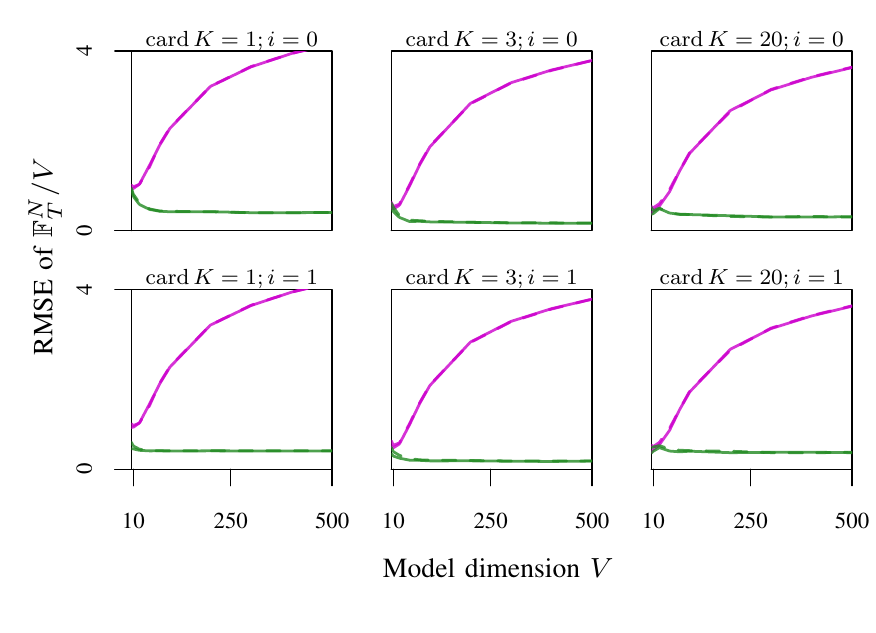}
    \caption{Using a \glsdesc{BPF} to approximate the filter.}\label{fig:rmse_bpf}
  \end{subfigure}
  \\
  \quad
  \begin{subfigure}[t]{\linewidth}
    \centering
  \includegraphics[trim=0.5cm 0cm 0cm 0.4cm]{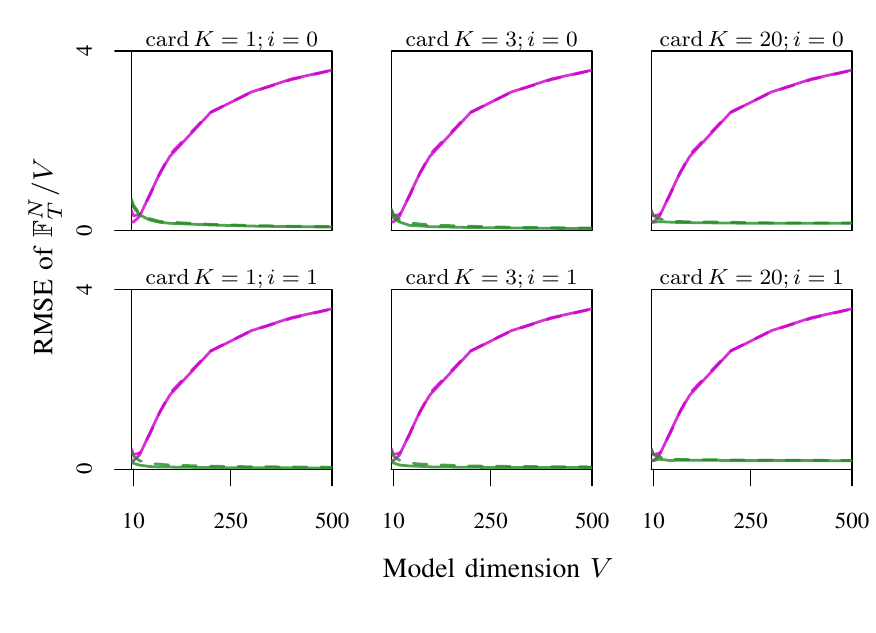}
    \caption{Using \gls{IID} samples from the filter.}\label{fig:rmse_mc}
  \end{subfigure}
  \caption{\Glsdesc{RMSE} of the estimate of $\smash{\smoothFunIntegral_\nTimeSteps/\modelDim}$. Obtained from $400$ simulation runs (each with a different observation sequence) using standard (\textcolor{plotmagenta}{\textbf{---}}) and blocked (\textcolor{plotgreen}{\textbf{---}}) \glsdesc{FS} as well as standard (\textcolor{plotmagenta}{\textbf{--~--}}) and blocked  (\textcolor{plotgreen}{\textbf{--~--}}) \glsdesc{BS}. Note that the lines for the standard smoothers are almost indistinguishable from one another and the same is true for the blocked smoothers.}\label{fig:rmse}
\end{figure}

We run standard and blocked \gls{FS} and \gls{BS} (with $\nParticles = 500$ and $\nBackwardPaths = 100$) for model dimensions up to $\modelDim = 500$, for contiguous blocks of size $\card \myBlock \in \{1,2,20\}$ and for enlarged blocks  $\enlargedBlock \coloneqq \neigh_i(\myBlock) = \bigcup_{v \in \myBlock}\neigh_i(v)$, for $i \in \{0,1\}$. The results in Fig.~\ref{fig:rmse} are based on $400$ independent repetitions, each using a different observation sequence $y_{1:20}$ sampled from the model. 

For each algorithm, we compare the impact of different filter approximations. Standard and blocked \glspl{PF} used the conditionally, locally optimal proposal kernel $\prop_{t,v}(x_{\neigh(v)}, z_v) \propto \trans_v(x_{\neigh(v)}, z_v) \obs(z_v, y_{t,v})$. While this proposal is usually intractable, in the model class considered in this work, it can be `exactly' approximated \citet{naesseth2015nested} (at an additional computational cost which grows in the model dimension). We employed the filter approximations from Assumptions~\ref{as:filter_approximation}\ref{as:filter_approximation:pf}--\ref{as:filter_approximation:pfIid} for the standard particle smoothers and from Assumptions~\ref{as:marginal_filter_approximation}\ref{as:marginal_filter_approximation:pf}--\ref{as:marginal_filter_approximation:pfIid} for the blocked particle smoothers. The results suggest the following interpretation.

\begin{myenumerate}[label=\textbullet]
 \item 
 Fig.~\ref{fig:rmse_smc} illustrates that the \gls{RMSE} of estimates of $\smash{\smoothFunIntegral_\nTimeSteps/\modelDim}$ grows in $\modelDim$ when using a standard \gls{PF} irrespective of the type of smoothing algorithm employed. This is consistent with our theory since Propositions~\ref{prop:asymptotic_variance} and \ref{prop:asymptotic_bias} assume dimensionally stable local filter approximations and standard \glspl{PF} (even with efficient proposals) normally break down in high dimensions. For instance, in a similar model, \citet{guarniero2016iterated} reported that even the \emph{fully-adapted auxiliary} \gls{PF} \citep{pitt1999filtering, johansen2008note} fails to yield useful filter approximations for model dimensions as small as $20$. 

 \item 
 Fig.~\ref{fig:rmse_bpf} and \ref{fig:rmse_mc} illustrate that when local filter errors are dimensionally stable, blocked particle smoothers yield estimates of $\smash{\smoothFunIntegral_\nTimeSteps/\modelDim}$ whose error is bounded in $\modelDim$. In contrast, standard particle smoothers induce an \gls{RMSE} that appears to grow with the model dimension.

 \item 
 The \gls{RMSE} of both standard and blocked particle smoothers is slightly higher in Fig.~\ref{fig:rmse_bpf} than in Fig.~\ref{fig:rmse_mc}. This is due to the additional bias induced by the \gls{BPF}.
 
 \item 
 The first and last column in Fig.~\ref{fig:rmse_bpf} and Fig.~\ref{fig:rmse_mc} illustrate the consequence of a suboptimal solution to the bias--variance trade-off discussed at the end of Subsection~\ref{subsec:theory}. That is, for the given number of particles, the block sizes $\card \myBlock = 1$ or $\card \myBlock = 20$ induce an error that is larger than for the choice $\card \myBlock = 3$ displayed in the second column.
\end{myenumerate}

\subsection{Parameter Estimation}
\label{subsec:parameter_estimation}

We now combine blocked particle smoothing with stochastic gradient-ascent and stochastic \gls{EM} algorithms in order to approximate the \gls{MLE} of the model parameters $\theta$ (see \citet{kantas2015particle} for a comprehensive review).

These algorithms generate a sequence of parameter values $\smash{(\thetaEstimate[p])_{p \in \naturals}}$ via  some update rule which requires the evaluation of some smoothed sufficient statistics $\smoothFunIntegral_\nTimeSteps^\theta$. As $\smoothFunIntegral_\nTimeSteps^\theta$ is usually intractable, we use (blocked) particle smoothers to estimate it. 

\subsubsection{Offline Gradient-ascent}  Let $(\gamma[p])_{p \in \naturals}$ be a step-size sequence which is non-negative, non-increasing, and which satisfies $\sum_{p =1}^\infty \gamma[p] = \infty$ as well as $\smash{\sum_{p =1}^\infty \gamma[p]^2 < \infty}$. Gradient-ascent algorithms use the update rule
\begin{equation}
 \thetaEstimate[p+1] = \thetaEstimate[p] + \gamma[p] \smoothFunIntegral_\nTimeSteps^{\thetaEstimate[p]}, \label{eq:offline_gradient_ascent}
\end{equation}
where $\smash{{\nabla_\vartheta \log \uSmooth_\nTimeSteps^\vartheta(\unitFun)|}_{\vartheta = \theta} = \smoothFunIntegral_\nTimeSteps^\theta = \smooth_\nTimeSteps^\theta(\smoothFun_\nTimeSteps^\theta)}$ is the score if the additive functionals $\testFun_{t,v}^\theta$ are as defined in Example~\ref{ex:score}.

Details on the calculation of the score for the model considered in this section are given in Subsection~I of the supplementary materials. 
The step sizes were $\smash{\gamma[p] = p^{-0.8}}$. To avoid manual tuning of the step-size sequence, we normalised the gradient approximation by its $L_2$ norm in \eqref{eq:offline_gradient_ascent}.

\subsubsection{Offline EM} \Gls{EM} algorithms use the update rule
\begin{equation}
 \thetaEstimate[p+1] \coloneqq \varLambda(\smoothFunIntegral_\nTimeSteps^{\thetaEstimate[p]}) \coloneqq \argmax_{\vartheta} \E\Bigl[\log \frac{\diff \uSmooth_\nTimeSteps^\vartheta}{\diff \refMeasX^{\otimes \nTimeSteps}}(X_{1:\nTimeSteps})\Bigr], \label{eq:offline_em}
\end{equation}
where $\smash{X_{1:\nTimeSteps} \sim \smooth_\nTimeSteps^\theta}$ and where the expectation on the \RHS is usually a function of $\smoothFunIntegral_\nTimeSteps^\theta$. 

The exact form of the function $\varLambda$ in \eqref{eq:offline_em} for the particular model considered in this section is also given in Subsection~I of the supplementary materials. 

\subsubsection{Results}

\begin{figure}[!t]
  \noindent{}
  \centering
  \includegraphics[trim=0.5cm 0.4cm 0cm 0.5cm]{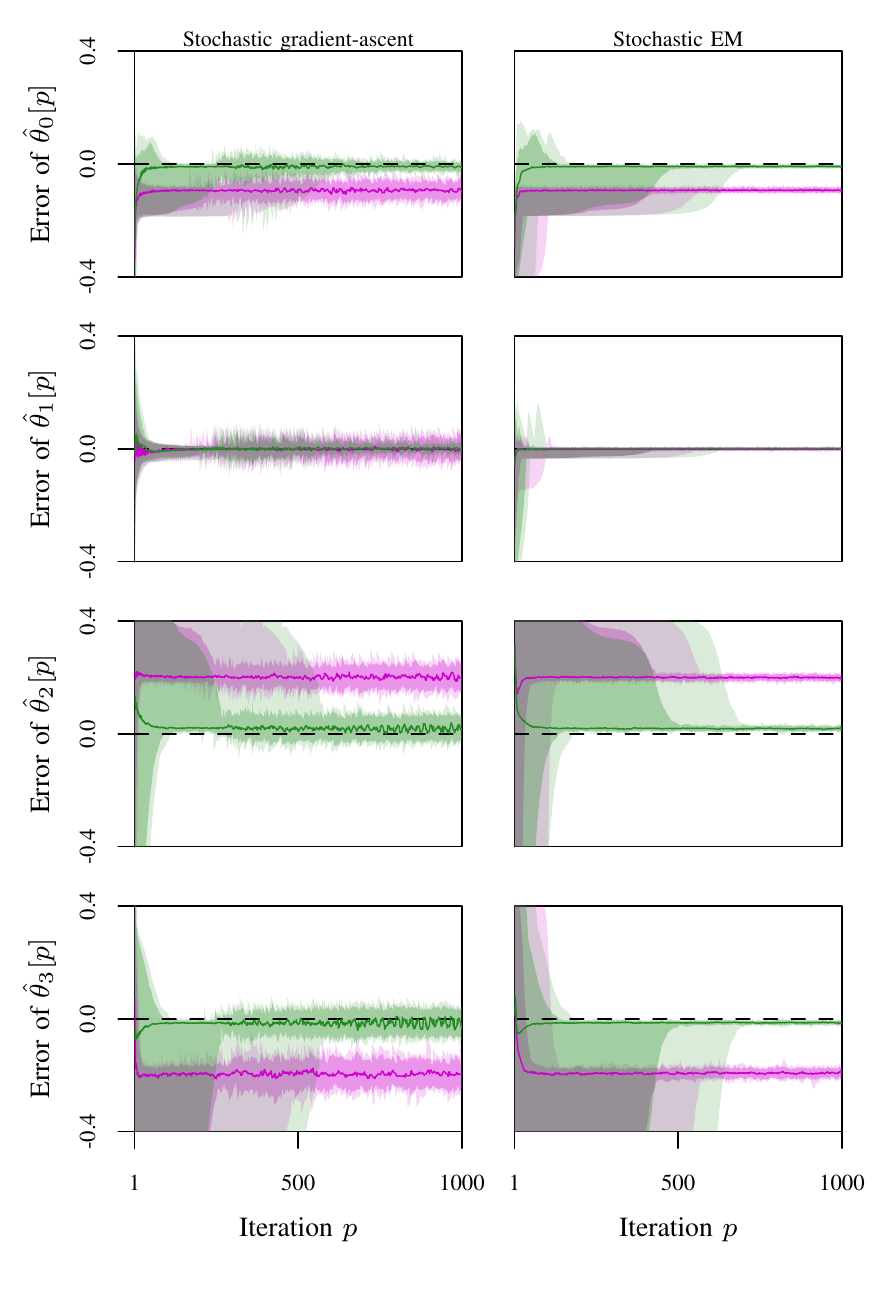}
  \caption{
  Average error of the parameter estimates in the $100$-dimensional state-space model. Obtained from $45$ runs of the stochastic gradient-ascent and stochastic \gls{EM} algorithms using standard (\textcolor{plotmagenta}{\textbf{---}}) and blocked (\textcolor{plotgreen}{\textbf{---}}) \glsdesc{BS}. We used blocks of size $\card \myBlock = 3$ and enlarged blocks of the form $\neigh_2(\myBlock)$. The shaded areas delimit, respectively, the range all encountered realisations and the $(0.05, 0.95)$-quantiles.
  }
  \label{fig:mle}	
\end{figure}

We apply both algorithms, based on standard and blocked \gls{BS}, to approximate the \gls{MLE} of $\theta$ using $\nParticles=500$, $\nBackwardPaths = 200$ and $\nTimeSteps =10$. In Fig.~\ref{fig:mle}, we only show results for filter approximations obtained from the \gls{BPF} (via Assumption~\ref{as:filter_approximation}\ref{as:filter_approximation:bpf} for standard \gls{BS} and via Assumption~\ref{as:marginal_filter_approximation}\ref{as:marginal_filter_approximation:bpf} for blocked \gls{BS}). As expected, using a standard \gls{PF} led to large errors in all algorithms.

%
%
%

\section{Summary}

We have presented online and offline smoothing recursions for efficiently estimating smoothed functionals in a class of high-dimensional state-space models (extensions to other models are discussed in Subsection~G of the supplementary materials). 
We combine these backward recursions with existing approximate forward-filtering recursions for high-dimensional models known as \glsdescplural{BPF}. The resulting algorithms then exploit spatial additivity of the smoothing functionals and spatial ergodicity of the model to produce estimates whose local errors are independent of the model dimension (for a fixed number of particles). Thus, we circumvent the so called curse of dimensionality. This is in contrast to existing methods which require the number of particles to scale exponentially in the model dimension. We have successfully applied our algorithms to perform smoothing and maximum-likelihood estimation.


\section*{Acknowledgements}
\addcontentsline{toc}{section}{Acknowledgements}

We wish to thank the four anonymous referees for numerous helpful comments and in particular for suggesting the discussion of ways in which our methodology may be extended to other types of state-space models (Subsection~G of the supplementary materials).

%


\appendix
\renewcommand{\theequation}{\thesection.\arabic{equation}}

\section*{Proof of Proposition~\ref{prop:exploding_asymptotic_variance}}
\label{app:proof_of_exploding_variance}

This appendix contains the proofs of Propositions~\ref{prop:exploding_asymptotic_variance}, \ref{prop:asymptotic_variance} and \ref{prop:asymptotic_bias}. To simplify the notation in all these proofs, we set $\smash{\obs_{t,v}(x_v) \coloneqq \obs_v(x_v, y_{t,v})}$, $\smash{\obs_{t,\myBlock}(x_\myBlock) \coloneqq \obs_\myBlock(x_\myBlock, y_{t,\myBlock})}$ and $\smash{\obs_{t}(x) \coloneqq \obs(x, y_{t})}$ for any $x \in \spaceX$, any $v \in \vertexSet$ and any $\myBlock \subseteq \vertexSet$.

Our proofs of Propositions~\ref{prop:exploding_asymptotic_variance} and \ref{prop:asymptotic_variance} are based on \citet[Theorem~3.1]{delmoral2010backward} which was formulated for prediction models but whose results can be easily transferred to updated models via \citet[Section~2.4.3]{delmoral2004feynman} as follows. The filter $\Filt_t$ of a state-space model defined by the triple $\modelOriginal$ can be interpreted as the predictor of a state-space model defined by the triple $(\breve{\init}, \breve{\trans}_t, \breve{\obs}_t)$, where 
  \begin{itemize}
    \item $\smash{\breve{m}(x) \coloneqq \init(z)\obs_1(z)/\int_\spaceX \init(z) \obs_1(z) \refMeasX(\diff z)}$, 
    \item $\smash{\breve{\trans}_t(x, z) \coloneqq \trans(x,z)\obs_t(z)/\int_\spaceX \trans(x, z)\obs_t(z) \refMeasX(\diff z)}$,
    \item $\smash{\breve{\obs}_t(x) \coloneqq \int_\spaceX \trans(x, z)\obs_{t+1}(z) \refMeasX(\diff z)}$.
  \end{itemize}

The remainder of this appendix is devoted to the proof of Proposition~\ref{prop:exploding_asymptotic_variance}.

\begin{proof}
   Let $\FiltComp_{t}$ and $\smoothComp_{\nTimeSteps}$ represent, respectively, the marginal of the time-$t$ filter and of the joint smoothing distribution on $\spaceXComp$, \IE by Assumption~\ref{as:spatially_iid_model}, $\smash{\Filt_t = \bigotimes_{v \in \vertexSet} \FiltComp_{t}}$ and $\smash{\smooth_\nTimeSteps = \bigotimes_{v \in \vertexSet}\smoothComp_{\nTimeSteps}}$. Let $\smash{\backComp_{\FiltComp_{t}}(\mathring{x}, \diff \mathring{z}) \coloneqq \FiltComp_{t}(\diff \mathring{z}) \transComp (\mathring{z}, \mathring{x}) / \FiltComp_{t}(\transComp(\ccdot, \mathring{x}))}$ be the backward kernel associated with the marginal of the model on one spatial component. Define $\QComp_s(\mathring{x}, \diff \mathring{z}) \coloneqq \transComp(\mathring{x}, \mathring{z}) \obsComp_s(\mathring{z}) \refMeasXComp(\diff \mathring{z})$,  $\smash{\QCalComp_{t,\nTimeSteps} \coloneqq \bigotimes_{s=t+1}^\nTimeSteps \QComp_s}$ as well as $G_{t,\nTimeSteps} \coloneqq \bigotimes_{v \in \vertexSet} \GComp_{t,\nTimeSteps}$, with $\GComp_{t,\nTimeSteps} \coloneqq {\QCalComp_{t,\nTimeSteps}(\unitFun)}/{\FiltComp_{t} \QCalComp_{t,\nTimeSteps}(\unitFun)}$ and, for $\mathring{z}_{t} \in \spaceXComp$, 
  \begin{align}
    \DComp_{t,\nTimeSteps}(\mathring{z}_{t}, \diff \mathring{x}_{1:\nTimeSteps})
    & \coloneqq \delta_{\mathring{z}_{t}}(\diff \mathring{x}_{t}) \Bigl[\smashoperator{\prod_{\smash{s=1}}^{t-1}} \backComp_{\FiltComp_{s}}(\mathring{x}_{s+1}, \diff \mathring{x}_{s})\Bigr]\!\!\!\\*
    & \qquad \times \QCalComp_{t,\nTimeSteps}(\mathring{x}_{t}, \diff \mathring{x}_{t+1:\nTimeSteps}),
  \end{align}
  Finally, $P_{t,\nTimeSteps} \coloneqq \bigotimes_{v \in \vertexSet} \PComp_{t,\nTimeSteps}$ with $\PComp_{t,\nTimeSteps} \coloneqq {\DComp_{t,\nTimeSteps}}/{\DComp_{t,\nTimeSteps}(\unitFun)}$.


%
  
  We can now prove Part~\ref{prop:exploding_asymptotic_variance:1}. By \citet[Theorem~3.1]{delmoral2010backward} -- transferring the results from prediction to updated models as outlined above -- the \gls{FS} approximation of $\smooth_\nTimeSteps(\smoothFun_\nTimeSteps)$ is asymptotically normal with asymptotic variance given by
  \begin{align}
  \sigma_\nTimeSteps^2(\smoothFun_\nTimeSteps) 
    & \coloneqq \sum_{t=1}^\nTimeSteps 
    \Filt_t\bigl(G_{t,\nTimeSteps}^2P_{t,\nTimeSteps}(\smoothFun_\nTimeSteps - \smooth_\nTimeSteps(\smoothFun_\nTimeSteps))^2\bigr). \label{eq:asymptotic_variance_sum}
  \end{align}
  Without loss of generality, assume that $\card J = 1$, \IE $J = \{u\}$, for some $u \in \vertexSet$. Write $\smoothFun_\nTimeSteps(x_{1:\nTimeSteps}) \equiv \smoothFunComp_\nTimeSteps(x_{1:\nTimeSteps,u}) \equiv \testFun_{r,u}(x_{r,u})$. The $t$th term on the \RHS of \eqref{eq:asymptotic_variance_sum} is then
  \begin{align}
    \MoveEqLeft\Filt_t\bigl(G_{t,\nTimeSteps}^2P_{t,\nTimeSteps}(\smoothFun_\nTimeSteps - \smooth_\nTimeSteps(\smoothFun_\nTimeSteps))^2\bigr)\\*
    & = \FiltComp_{t}\bigl( \GComp_{t,\nTimeSteps}^2\PComp_{t,\nTimeSteps}(\smoothFunComp_{\nTimeSteps} - \smoothComp_\nTimeSteps(\smoothFunComp_{\nTimeSteps}))^2\bigr) \FiltComp_{t}(\GComp_{t,\nTimeSteps}^2)^{\modelDim - 1}\!\!\!\!\!\!\!\!\!.\qquad\label{eq:proof_eq_1}
  \end{align}
  Since Jensen's inequality ensures that
  \begin{align}
    c_{t, \nTimeSteps} & \coloneqq \FiltComp_t(\GComp_{t,\nTimeSteps}^2) \geq \FiltComp_t(\GComp_{t,\nTimeSteps})^2 = 1, \label{eq:proof_strict_inequality}
  \end{align}
  and since $\testFun_{r,J}$ is not $\refMeasXComp$-almost everywhere constant, 
  \begin{align}
    a_{t, \nTimeSteps}(\testFun_{r,J})& \coloneqq \FiltComp_t\bigl(\GComp_{t,\nTimeSteps}^2\PComp_{t,\nTimeSteps}(\smoothFunComp_\nTimeSteps - \smoothComp_\nTimeSteps(\smoothFunComp_\nTimeSteps))^2\bigr) /c_{t, \nTimeSteps} > 0.
  \end{align}
  This proves Part~\ref{prop:exploding_asymptotic_variance:1}.

  It remains to prove Part~\ref{prop:exploding_asymptotic_variance:2}. If the model transitions are not perfectly mixing, $\smash{\GComp_{t,\nTimeSteps}}$ is not $\refMeasXComp$-almost everywhere constant. Hence, Jensen's inequality in \eqref{eq:proof_strict_inequality} is strict (since $x \mapsto x^2$ is strictly convex). As a result, the \RHS in \eqref{eq:proof_eq_1}, and thus $\sigma_\nTimeSteps^2(\smoothFun_\nTimeSteps)$, grows exponentially in $\modelDim$. \qedhere
\end{proof}

\section*{Proofs for Subsection~\ref{subsec:theory}} 
\label{app:proofs_of_variance_and_bias_bounds}

\subsection{Outline}
In this section, we prove Propositions~\ref{prop:asymptotic_variance} and \ref{prop:asymptotic_bias}. Central to the proofs are three alternate state-space models which may be viewed as approximations of the original state-space model, \IE approximations of the model defined by the triple $\modelOriginal$.
\begin{myenumerate}
 \item Model $\modelTilde$, defined in Subsection~\ref{subsec:model_tilde}, represents the asymptotic target distribution of the \gls{BPF} as stated in \citet{rebeschini2014comparison}. It defines a joint smoothing distribution $\smoothTilde_\nTimeSteps$ and filters $\FiltTilde_t$.
 \item Model $\modelHat$, defined in Subsection~\ref{subsec:model_hat}, defines a joint smoothing distribution $\smoothHat_\nTimeSteps$ whose marginal on Block~$\myBlock \supseteq J$ coincides with the corresponding marginal of $\smooth_\nTimeSteps$ and its filter coincides with $\Filt_t$.
  \item Model $\modelBar$, defined in Subsection~\ref{subsec:model_bar}, defines a joint smoothing distribution $\smoothBar_\nTimeSteps$ whose marginal on Block~$\myBlock \supseteq J$ coincides with the corresponding marginal of $\smoothTilde_\nTimeSteps$ and its filter coincides with $\FiltTilde_t$.
\end{myenumerate}

When approximating the filter using \gls{IID} samples from $\Filt_t$ (respectively $\FiltTilde_t$), blocked particle smoothing for $\modelOriginal$ coincides with standard particle smoothing for $\modelHat$ (respectively $\modelBar$). The central limit theorem for standard particle smoothing from \citet{delmoral2010backward} then proves Proposition~\ref{prop:asymptotic_variance}.

When approximating the filter using the \gls{BPF} (or \gls{IID} samples from $\FiltTilde_t$), blocked particle smoothing for $\modelOriginal$ coincides with standard particle smoothing for $\modelTilde$. The bound on the bias from \citet{rebeschini2015local} then proves Proposition~\ref{prop:asymptotic_bias}. 

For probability measures $\mu$ and $\nu$ on $\mathbb{S} = \prod_{i \in \mathbb{I}} \mathbb{S}_i$ and $I \subseteq \mathbb{I} \subseteq \naturals$, define the \emph{local total variation distance} 
\begin{equation}
 \lVert \mu - \nu \rVert_I \coloneqq \sup_{\testFun \in \calS_I} \lvert \mu(\testFun) - \nu(\testFun)\rvert,
\end{equation}
where $\calS_I$ is the class of measurable functions $\testFun\colon \mathbb{S} \to [-1,1]$ such that for all $x,z \in \mathbb{S}$, $x_I = z_I$ implies $\testFun(x) = \testFun(z)$.

\subsection{Model $\modelTilde$}
\label{subsec:model_tilde}


The state-space model $\modelTilde$ is defined by $\initTilde \coloneqq \init$, $\obsTilde_t \coloneqq \obs_t$, and $\transTilde_t(x,z) \coloneqq \prod_{\myBlock \in \myBlockSet} \transTilde_{t,\myBlock}(x_\myBlock, z_\myBlock)$ with
\begin{align}
\!\!\!\!\!
 \transTilde_{t,\myBlock}(x_{\myBlock}, z_{\myBlock})
 & \coloneqq \int_{\spaceX} \!\trans_\myBlock(x_{\neigh(\myBlock)}, z_\myBlock) \FiltTilde_{t-1,\neigh(\myBlock)\setminus \myBlock}(\diff x_{\neigh(\myBlock)\setminus \myBlock}).\!\!\!\!\!
\end{align}
Let $\smoothTilde_t$ be the joint smoothing distribution of this model (on $\spaceX^t$) and $\FiltTilde_t(A) \coloneqq \smoothTilde_t(\unitFun \otimes \ind_A)$, for $A \subseteq \spaceX$ and $t > 1$, is the blocked filter, with initial condition $\FiltTilde_1 \coloneqq \smoothTilde_1 = \Filt_1$. 

\subsection{Model $\modelHat$}
\label{subsec:model_hat}

The state-space model $\modelHat$ is defined  by 
\begin{align}
 \initHat(z) & \coloneqq \init_\myBlock(z_\myBlock){\filt_1(z)}/{\filt_{1,\myBlock}(z_\myBlock)},\\*
 \transHat_{t}(x, z)
 & \coloneqq \trans_{\myBlock}(x_{\neigh(\myBlock)}, z_{\myBlock}) 
 {\filt_t(z)}/{\filt_{t,\myBlock}(z_\myBlock)}, \quad \text{for $t>1$,}\\*
 \obsHat_t(x) & \coloneqq \obs_{t,\myBlock}(x_{\myBlock}).
\end{align}
Here, writing $\smash{\myBlock^\compl \coloneqq \vertexSet \setminus \myBlock}$ we have defined $\smash{\filt_t(x) \coloneqq \frac{\diff \Filt_{t}}{\diff \refMeasX}(x)}$ and $\smash{\filt_{t,\myBlock}(x_\myBlock) \coloneqq \int_{\spaceX_{\myBlock^\compl}} \filt_t(x) \refMeasX_{\myBlock^\compl}(\diff x_{\myBlock^\compl})}$.
We let $\smoothHat_t$, $\FiltHat_t$ and $\backHat_{t, \FiltHat_t}(x, \diff z) \coloneqq {\transHat_{t+1}(z, x)\FiltHat_t(\diff z) }/{\FiltHat_t(\transHat_{t+1}(\ccdot, x))}$ be the joint smoothing distribution, filter and standard backward kernel of $\modelHat$.

\begin{lemma}\label{lem:properties_of_modelHat}~
 \begin{myenumerate}
  \item \label{lem:properties_of_modelHat:1} For any $t \in \naturals$, $\lVert \FiltHat_{t} - \Filt_t \rVert_{\vertexSet} = 0$.
  \item \label{lem:properties_of_modelHat:2} For any $t \in \naturals$, $  \lVert \smoothHat_{t} - \smooth_{t}\rVert_{\{1,\dotsc,t\} \times \myBlock}= 0$. 
  \item \label{lem:properties_of_modelHat:3} Under Assumption~\ref{as:local_test_function}, 
  \begin{align}
  \smoothHat_{\nTimeSteps}(\smoothFun_\nTimeSteps)
  & = \int \testFun_{r,J}(x_{r,J})\Filt_{\nTimeSteps,\myBlock}(\diff x_{\nTimeSteps,\myBlock})\\*
  & \quad \times \prod_{t=1}^{\smash{\nTimeSteps-1}} \back_{\myBlock, \Filt_{t,\neigh(\myBlock)}}(x_{t+1,\myBlock}, \diff x_{t,\neigh(\myBlock)}),
  \end{align}
  where $\smash{\back_{\myBlock, \nu}}$ is the blocked backward kernel from \eqref{eq:blocked_backward_kernel}
 \end{myenumerate} 
\end{lemma}

The proof of Parts~\ref{lem:properties_of_modelHat:1} and \ref{lem:properties_of_modelHat:2} of Lemma~\ref{lem:properties_of_modelHat} follow by induction. Part~\ref{lem:properties_of_modelHat:3} follows from Part~\ref{lem:properties_of_modelHat:1} and the definition of $\backHat_{t, \Filt_t}$. For completeness, we detail these proofs in Subsection~H of the supplementary materials. 


\subsection{Model $(\bar{\init}, \transBar_t, \obsBar_t)$}
\label{subsec:model_bar}

The state-space model $\modelBar$ is defined by 
\begin{align}
 \initBar(z) & \coloneqq \init_\myBlock(z_\myBlock) \filtTilde_{1,\myBlock^\compl}(z_{\myBlock^\compl}), \label{eq:initBar}\\*
 \transBar_{t}(x, z)
 & \coloneqq \trans_{\myBlock}(x_{\neigh(\myBlock)}, z_{\myBlock}) \filtTilde_{t,\myBlock^\compl}(z_{\myBlock^\compl}), \quad \text{for $t>1$,} \label{eq:transBar}\\
 \obsBar_t(x) &\coloneqq \obs_{t,\myBlock}(x_{\myBlock}).
\end{align}
Here, for any $\myBlock \subseteq \vertexSet$, we have defined
$\smash{\filtTilde_t(x) \coloneqq \frac{\diff \FiltTilde_{t}}{\diff \refMeasX}(x)}$ and $\smash{\filtTilde_{t,\myBlock^\compl}(x_{\myBlock^\compl}) \coloneqq \int_{\spaceX_{\myBlock}} \filtTilde_t(x) \refMeasX_{\myBlock}(\diff x_{\myBlock})}$. We let $\smoothBar_t$, $\FiltBar_t$ and $\backBar_{t, \FiltBar_t}(x, \diff z) \coloneqq {\transBar_{t+1}(z, x)\FiltBar_t(\diff z) }/{\FiltBar_t(\transBar_{t+1}(\ccdot, x))}$ be the joint smoothing distribution, filter and standard backward kernel of $\modelBar$. 

\begin{lemma}\label{lem:properties_of_modelBar}~
 \begin{myenumerate}
  \item \label{lem:properties_of_modelBar:1} For any $t \in \naturals$, $\lVert \FiltBar_{t} - \FiltTilde_t \rVert_{\vertexSet} = 0$.
  \item \label{lem:properties_of_modelBar:2} For any $t \in \naturals$, $\lVert \smoothBar_{t} - \smoothTilde_{t}\rVert_{\{1,\dotsc,t\} \times \myBlock}= 0$. 
  \item \label{lem:properties_of_modelBar:3} Under Assumption~\ref{as:local_test_function}, 
  \begin{align}
  \smoothBar_{\nTimeSteps}(\smoothFun_\nTimeSteps)
  & = 
  \int \testFun_{r,J}(x_{r,J})\FiltTilde_{\nTimeSteps,\myBlock}(\diff x_{\nTimeSteps,\myBlock})\\* & \quad \times \prod_{t=1}^{\smash{\nTimeSteps-1}} \back_{\myBlock, \FiltTilde_{t,\neigh(\myBlock)}}(x_{t+1,\myBlock}, \diff x_{t,\neigh(\myBlock)}),
  \end{align}
  where $\smash{\back_{\myBlock, \nu}}$ is the blocked backward kernel from \eqref{eq:blocked_backward_kernel}.
 \end{myenumerate}
\end{lemma}

The proof of Lemma~\ref{lem:properties_of_modelBar} is similar to the proof of Lemma~\ref{lem:properties_of_modelHat}. For completeness, we detail the proof in Subsection~H of the supplementary materials.

\subsection{Central Limit Theorem}



\begin{proof}[of Proposition~\ref{prop:asymptotic_variance}] We first prove Part~\ref{prop:asymptotic_variance:2}. By Lemma~\ref{lem:properties_of_modelHat}, we are performing standard particle smoothing for $\modelHat$. For any $t \leq q \leq \nTimeSteps$ and any $x_t, z_t \in \spaceX$, define
\begin{align}
 \mathcal{Q}_{t,q}(x_t, \diff x_{t+1:\nTimeSteps}) 
 & \coloneqq \smashoperator{\prod_{s=t+1}^q} \transHat_s(x_{s-1}, x_s) \obsHat_s(x_s) \refMeasX(\diff x_s),\\
 D_{t,\nTimeSteps}(z_t, \diff x_{1:\nTimeSteps}) 
 & \coloneqq \updelta_{z_t}(\diff x_t) \biggl[\prod_{\smash{s=1}}^{\smash{t-1}} \backHat_{s, \FiltHat_{s}}(x_{s+1}, \diff x_s)\Bigr]\\*
  & \qquad \times \mathcal{Q}_{t,\nTimeSteps}(x_t, \diff x_{t+1:\nTimeSteps}).
\end{align}
Furthermore, we write $Q_{t,q}(f_q)(x_t) \coloneqq \mathcal{Q}_{t,q}(\unitFun \otimes f_q)(x_t)$, and also define $G_{t,\nTimeSteps} \coloneqq {Q_{t,\nTimeSteps}(\unitFun)}/{\FiltHat_t Q_{t,\nTimeSteps}(\unitFun)}$ as well as $P_{t,\nTimeSteps}(\smoothFun_\nTimeSteps) \coloneqq {D_{t,\nTimeSteps}(\smoothFun_\nTimeSteps)}/{D_{t,\nTimeSteps}(\unitFun)}$.

By \citet[Theorem~3.1]{delmoral2010backward} (again transferred to updated models as described above) the blocked \gls{FS} approximation of $\smoothHat_{\nTimeSteps}(\smoothFun_{\nTimeSteps})$ is asymptotically normal with asymptotic variance
\begin{align}
 \sigma_{\nTimeSteps}^2(\smoothFun_{\nTimeSteps}) 
 & \leq \varepsilon^{4 \card \myBlock} \sum_{t=1}^{\smash{\nTimeSteps}} \FiltHat_t\bigl(P_{t,\nTimeSteps}(\smoothFun_{\nTimeSteps} - \smoothHat_{\nTimeSteps}(\smoothFun_{\nTimeSteps}))^2\bigr),
\end{align}
since it is easy to check that $\smash{G_{t,\nTimeSteps} \leq \varepsilon^{2\card \myBlock}}$. 
Define the Markov kernels $R_{t,r}\colon \spaceX \times \Borel(\spaceX) \to [0,1]$ by
\begin{align}
 \MoveEqLeft R_{t,r}(\testFun_r)(x_t)\\*
 & \!\!\!\!\!\!\!\!\coloneqq 
 \begin{cases}
  \delta_{x_t}(\testFun_r) = \testFun_r(x_t), & \text{if $t = r$,}\\
  Q_{t,r}(\testFun_r)(x_t)/Q_{t,r}(\unitFun)(x_t), & \text{if $t < r$,}\\
  \int_{\spaceX^{t-r}} \testFun_r(x_r)\prod_{s=t-1}^r \backHat_{s,\FiltHat_s}(x_{s+1}, \diff x_s), & \text{if $t > r$.}
 \end{cases}
\end{align}
Let $\osc(\testFun) \coloneqq \sup_{(x,y)\in\spaceX^2} \lvert \testFun(x)-\testFun(y)\rvert$ denote the oscillations of some  function $\testFun\colon \spaceX \to \reals$ and let $\Osc_1(\spaceX)$ denote the set of all real-valued  functions with domain $\spaceX$ whose oscillations are less than $1$. Furthermore, let  $\beta(M) \coloneqq \sup_{\testFun \in \Osc_1(\spaceX)} \osc(M(\testFun)) \in [0,1]$ denote the Dobrushin coefficient of a Markov kernel $M\colon \spaceX \times \Borel(\spaceX) \to [0,1]$. Then
\begin{align}
 \FiltHat_t\bigl(P_{t,\nTimeSteps}(\smoothFun_{\nTimeSteps} - \smoothHat_{\nTimeSteps}(\smoothFun_{\nTimeSteps}))^2\bigr) 
 \leq 4 \beta(R_{t,r}), \label{eq:dobrushin_as_bound}
\end{align}
where we have used that since $\lVert \testFun_{r,J}\rVert \leq 1$, 
\begin{equation}
 \osc\bigl((\smoothFun_{\nTimeSteps} - \smoothHat_{\nTimeSteps}(\smoothFun_{\nTimeSteps}) )/2\bigr) = \osc\bigl((\testFun_{r,J} - \smoothHat_{\nTimeSteps}(\smoothFun_{\nTimeSteps}) )/2\bigr)\leq 1.
\end{equation}
For any $t \in \naturals$, we bound $\beta(R_{t,r})$ in \eqref{eq:dobrushin_as_bound} as follows:
\begin{equation}
 \beta(R_{t,r}) \leq (1 - \varepsilon^{-4\card \myBlock})^{\lvert t-r\rvert}. \label{eq:dobrushin_R_inequality}
\end{equation}
If $t=r$, \eqref{eq:dobrushin_R_inequality} holds because $\beta(R_{t,r}) \leq 1$. If $t<r$, \eqref{eq:dobrushin_R_inequality} is implied by \citet[Proposition~4.3.3]{delmoral2010backward}. Finally, it can be easily checked that $\backHat_{t,\FiltHat_t}(x, \ccdot) \geq \varepsilon^{-4 \card \myBlock}\backHat_{t,\FiltHat_t}(z, \ccdot)$, for  any $(x,z) \in \spaceX^2$, so that $\beta(\backHat_{t,\FiltHat_t}) \leq 1 - \varepsilon^{-4 \card \myBlock}$. Hence, if $t>r$, $\beta(R_{t,r}) \leq \prod_{s=t-1}^r \beta(\backHat_{s,\FiltHat_s}) \leq (1 - \varepsilon^{-4\card \myBlock})^{\lvert t-r\rvert}$. As a result,
\begin{align}
 \sum_{t=1}^\nTimeSteps \beta(R_{t,r}) \leq 2 \sum_{t=0}^\infty (1 - \varepsilon^{-4 \card \myBlock})^t = 2 \varepsilon^{4 \card \myBlock},
\end{align}
so that $\smash{\sigma_{\nTimeSteps}^2(\smoothFun_{\nTimeSteps}) 
\leq  8 \varepsilon^{8 \card \myBlock} = \eul^{\consta \card \myBlock} \leq \eul^{\consta \maxBlockSize}}$, with $\consta \coloneqq \log(8 \varepsilon^{8})$. This completes the proof of Part~\ref{prop:asymptotic_variance:2}.

To prove Part~\ref{prop:asymptotic_variance:1}, note that by Lemma~\ref{lem:properties_of_modelBar}, we are performing standard particle smoothing for $\modelBar$. Part~\ref{prop:asymptotic_variance:2} is then proved exactly as Part~\ref{prop:asymptotic_variance:1} but with $(\initHat, \transHat_t, \obsHat_t, \smoothHat_\nTimeSteps, \FiltHat_t, \backHat_{t, \FiltHat_t})$ replaced by $(\initBar, \transBar_t, \obsBar_t, \smoothBar_\nTimeSteps, \FiltBar_t, \backBar_{t, \FiltBar_t})$. \qedhere
\end{proof}

\subsection{Asymptotic Bias}


\begin{proof}[of Proposition~\ref{prop:asymptotic_bias}]
 By Lemma~\ref{lem:properties_of_modelBar} we are performing standard particle smoothing for $\modelTilde$. In particular, 
 \begin{equation}
  \lVert \smoothBar_{\nTimeSteps} - \smooth_\nTimeSteps \rVert_{\{t\} \times J} = \lVert \smoothTilde_\nTimeSteps  - \smooth_\nTimeSteps \rVert_{\{t\} \times J},
 \end{equation}
 for any $t \in \{1,\dotsc, \nTimeSteps\}$ and any $J \subseteq \myBlock$. Then by \citet[Theorem~4.3]{rebeschini2014comparison} (which is stated only for the bias of the filter, \IE for $t=\nTimeSteps$, but whose proof is established for $t<\nTimeSteps$),
\begin{align}
 \lVert \smoothTilde_\nTimeSteps  - \smooth_\nTimeSteps \rVert_{\{t\} \times J}
 & \leq \constb \card(J) \eul^{-\constc \metric(J, \partial \myBlock)}.
\end{align}
 This completes the proof. \qedhere
\end{proof}

\bibliographystyle{IEEEtran}
\bibliography{block}

%
\begin{IEEEbiography}[{\includegraphics[width=1in,height=1.25in,clip,keepaspectratio]{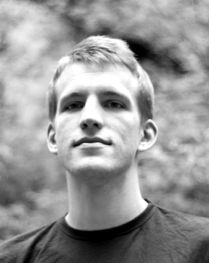}}]{Axel~Finke}
 received a B.S.\ (2010) in economics from M\"unster Univ., Germany and an M.S.\ (with distinction) and Ph.D.\ in statistics from Warwick Univ., UK in 2011 and 2015, respectively. Subsequently, he joined the Signal Processing and Communications Laboratory, Dept.\ of Engineering, Univ.\ of Cambridge, UK, as research associate. He is currently working as a research associate at the Dept.\ of Statistical Science, University College London, UK His research interests include computational statistics, in particular sequential Monte Carlo and Markov chain Monte Carlo methods for both Bayesian and Frequentist inference.
\end{IEEEbiography}

\begin{IEEEbiography}[{\includegraphics[width=1in,height=1.25in,clip,keepaspectratio]{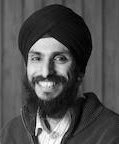}}]{Sumeetpal~S.~Singh}
 received the B.E.\ (with first-class honours) and Ph.D.\ degrees from the Dept.\ of Electrical Engineering, University of Melbourne, Australia, in 1997 and 2002, respectively. After having worked in Industry for a number of years, he joined the Cambridge University Engineering Department in 2004 as a Research Associate and is currently a Reader in Engineering Statistics. He is also a Fellow and Director of Studies at Churchill College, a Fellow of the Alan Turing Institute, an Affiliated Lecturer of the Statistics Laboratory and an Associate Editor of Statistics and Computing. His research focus is on statistical signal processing, in particular, using Monte Carlo methods, covering algorithmic development for applications and theoretical analysis.  He has been recognised for his work on Multi-target Tracking (awarded the IEEE M.~Barry Carlton Award in 2013.)
\end{IEEEbiography}


\clearpage
\pagenumbering{roman}
\setcounter{page}{1}

%


\renewcommand{\theequation}{\thesection.\arabic{equation}}

\newgeometry{left=5cm, right=5cm, top=2cm, bottom=3cm}
\onecolumn

\begin{center}
 \Large{Supplementary Materials:\\\emph{Approximate Smoothing and Parameter Estimation in High-Dimensional State-Space Models}}\\ \vspace{2ex}
 \normalsize{Axel Finke, Sumeetpal S.~Singh}
 \vspace{2ex}
\end{center}

In these supplementary materials, we first discuss ways in which the methods proposed in the main document may be extended to other models than those described in Subsection~\ref{subsec:high-dimensional_ssm} in the main document. We also provide the proofs of Lemmata~\ref{lem:properties_of_modelHat} and \ref{lem:properties_of_modelBar} as well as additional details about the simulation study conducted in Section~\ref{sec:simulations} of the main document. 

\subsection{Extension to other models}
\label{subsec:extension_to_other_models}

    To develop dimensionally stable particle smoothers, it is necessary to exploit some specific model structure. Indeed, as in the case of particle filters, it seems unlikely that there exists a way of circumventing the curse of dimensionality which is directly applicable to \emph{arbitrary} state-space models. 
    
    In this work, we exploit such structure by considering the canonical class of high-dimensional state-space models from \citet{rebeschini2015localAlt} (reviewed in Section~\ref{subsec:high-dimensional_ssm} of the main document). In this class of models, we can exploit the spatial-decorrelation to circumvent the curse of dimensionality suffered by particle \emph{smoothers} via suitable blocking approximations (\IE using the idea through which \citet{rebeschini2015localAlt} devised dimensionally stable particle \emph{filters}). 
    
    Of course, many existing high-dimensional state-space models do not exactly fit into the class of models from Subsection~\ref{subsec:high-dimensional_ssm} in the main document (hereafter, we will simply refer to the latter as `canonical models') but will nonetheless exhibit (or can be modified to exhibit) some spatial decorrelation. To perform smoothing for such models, we propose two options.
    
    \begin{myenumerate}
      \item \label{enum:modified_algorithm} In some cases, it may be possible to devise blocked particle methods such that they still exploit spatial decorrelation in the spirit of \citet{rebeschini2015localAlt} but can be implemented for the original model. This approach is discussed in Example~\ref{ex:choice_of_approximate_algorithm} below, where we describe a way of modifying our blocked particle smoothers to accommodate a different class of model.
      
      \item \label{enum:modified_model} Alternatively,  we advocate taking the bias--variance trade-off which is already at the heart of blocking approximations (see \citet{rebeschini2015localAlt} and Subsection~\ref{subsec:theory} of the main document) one step further. That is, we advocate accepting slightly more model misspecification error (\IE bias) in exchange for a practical (\IE dimensionally stable) way of estimating the model. Example~\ref{ex:choice_of_approximate_model} below describes how learning two more parameters (one of which is a suitable neighbourhood size $\neighSize$) for the canonical model allows us to relax the assumption that the state transition of the fitted model fully factorise as specified in Subsection~\ref{subsec:high-dimensional_ssm} of the main document. Our proposed methodology and analysis applies without modification in this case.
    \end{myenumerate}

  \begin{example}[modified algorithm] \label{ex:choice_of_approximate_algorithm} As before, let $X_t$ and $Y_t$ be $\modelDim$-dimensional vectors. Consider the multivariate stochastic volatility model defined through
    \begin{align}
      Y_{t} & = V_t \eta_{t},\\
      X_{t} & =AX_{t-1}+\epsilon_{t},
    \end{align}
    where $V_t$ is a $\modelDim \times \modelDim$ diagonal matrix whose entries on the diagonal are given by the volatilities $\exp(X_{t,1}), \dotsc, \exp(X_{t,\modelDim})$. The joint error term $(\epsilon_{t}, \eta_{t})$ is multivariate Gaussian with covariance matrix
    \begin{equation}
    \begin{bmatrix}
     \varSigma_\epsilon & \varSigma_{\epsilon, \eta}\\
     \varSigma_{\epsilon, \eta} &   \varSigma_\eta
    \end{bmatrix}.
    \end{equation}
    We assume that order of the assets, \IE the order of the components of $X_t$ and $Y_t$, has been chosen in such a way that highly correlated assets have been grouped together (\EG by applying some clustering technique to the data). This ensures spatial decorrelation by implying that $A$, $\varSigma_\epsilon$ and $\varSigma_\eta$ have little mass far away from the main diagonal. Note, however, that we do not assume that the entries off the main diagonal (or off the first few diagonals in the case of $A$) are zero so that the transition densities $\trans(x,z)$ and $\obs(z, y_t)$ (with respect to the usual dominating measure $\refMeasX = \bigotimes_{v\in \vertexSet} \refMeasX_v$ and $\refMeasY = \bigotimes_{v\in \vertexSet} \refMeasY_v$ for $\vertexSet \coloneqq \{1, \dotsc, \modelDim\}$) do not factorise. As a result, the stochastic volatility model cannot be viewed as a special case of the canonical model. 
    
    The model does, however, satisfy the following weakened assumption. 
    
    \begin{assumption}\label{as:msv}
     For any time~$t$ and any block $\myBlock \subseteq \vertexSet$, it is possible to evaluate the marginal densities
    \begin{myenumerate}[label=\alph*), ref=\alph*]
      \item \label{as:msv:filter} $\obs_\myBlock(z,y_{t,\myBlock}) \coloneqq \int \obs(z,y_{t})\prod_{v\notin \myBlock}\refMeasY_v(\diff y_{t,v})$,
      \item \label{as:msv:smoother} $\trans_\myBlock(x,z_{K}) \coloneqq \int p(x,z)\prod_{v\notin \myBlock}\refMeasX_v(\diff z_{v})$.
    \end{myenumerate}
    \end{assumption}

    Under Assumption~\ref{as:msv}\ref{as:msv:filter}, we can approximate the filter via a modified version of the \gls{BPF} (for simplicity, we only consider a bootstrap version, \IE $\prop_t(x, z) = \trans(x,z)$). The modified \gls{BPF} remains exactly as in Algorithm~\ref{alg:blocked_pf} of the main document except that the local weights associated with block $\myBlock$ are now calculated as
    \begin{align}
     w_{t,\myBlock}^n \coloneqq \obs_\myBlock(x_{t}^n, y_{t,\myBlock}). \label{eq:modified_weight}
    \end{align}

    Likewise, under Assumption~\ref{as:msv}\ref{as:msv:smoother}, we can perform smoothing via modified versions of blocked \gls{FS} and blocked \gls{BS}. The modified blocked particle smoothers remain exactly as Algorithms~\ref{alg:blocked_fs} and \ref{alg:blocked_bs} of the main document, except that we replace the blocked backward kernels $\smash{\back_{\myBlock, \Filt^\nParticles_{t,\neigh(\myBlock)}}(x_\myBlock, \diff z_{\neigh(\myBlock)})}$ by
    \begin{align}
      \sum_{n=1}^\nParticles \frac{W_{t,\myBlock}^n \trans(X_t^n, x_\myBlock)}{\sum_{k = 1}^\nParticles W_{t,\myBlock}^k \trans(X_t^k, x_\myBlock)} \delta_{X_{t,\myBlock}^n}(\diff z_\myBlock). \label{eq:modified_backward_kernel}
    \end{align}
    Note that these kernels now map from $\myBlock$ to $\myBlock$ (in a suitable sense) whereas Algorithms~\ref{alg:blocked_fs} and \ref{alg:blocked_bs} (main document) used kernels that map from $\myBlock$ to $\neigh(\myBlock)$. Though if required by the test functions which we wish to integrate, we could easily make \eqref{eq:modified_backward_kernel} map from $\myBlock$ to some $i$-neighbourhood $\neigh_i(\myBlock)$ by replacing $\smash{X_{t,\myBlock}^n}$ and $\smash{W_{t,\myBlock}^n}$ by $\smash{X_{t,\neigh_i(\myBlock)}^n}$ and $\smash{W_{t,\neigh_i(\myBlock)}^n}$ (computed via \eqref{eq:modified_weight}) in \eqref{eq:modified_backward_kernel}. As usual, we may also replace $\myBlock \in \myBlockSet$ by some enlarged block $\enlargedBlock \supseteq \myBlock$ to reduce bias.
  \end{example}
  
   \begin{example}[modified model] \label{ex:choice_of_approximate_model}
      Note that even though the canonical model assumes that spatial components interact only locally (in some $\neighSize$-neighbourhood) over a single time step, all spatial components interact with one another after sufficiently many time steps $L$, where $L$ is a function of the neighbourhood size $\neighSize$. 
      
      For instance, in the one-dimensional lattice example shown in Fig.~1 -- where $\neighSize =1$, \IE $\neigh(v) = \{v-1, v, v+1\}$ -- all spatial components interact over $L = \modelDim-1$ steps. 
      
      This motivates approximating a model with non-local spatial interactions (and observations $y_{1:\nTimeSteps'}$ over $\nTimeSteps'$ time steps) by a canonical model with only local interactions and with $\nTimeSteps \coloneqq L \nTimeSteps'$ and where the latter takes 
    \begin{align}
      \obs(x_t, y_t) \coloneqq
     \begin{cases}
      \obs(x_t, y_t'), & \text{for $t \in \{L, 2L, \dotsc, L\nTimeSteps'\}$,}\\
      1, & \text{otherwise.}
     \end{cases}
    \end{align}
     Given the choice of approximating canonical model, our theoretical results immediately guarantee dimensional stability of the blocked particle smoothers as the latter can be applied without any change. In particular, the parameter-estimation algorithms from Subsection~\ref{subsec:parameter_estimation} of the main document can then be used to guide the selection of $\neighSize$ and $L$.
   \end{example}

\subsection{Additional proofs}
\label{subsec:additional_proofs}

In this subsection, we present the proofs of Lemma~\ref{lem:properties_of_modelHat} and Lemma~\ref{lem:properties_of_modelBar} which were omitted from the main document.

\begin{proof}[of Lemma~\ref{lem:properties_of_modelHat}]
 We prove Part~\ref{lem:properties_of_modelHat:1} by induction. Clearly,
 \begin{align}
  \FiltHat_1(\diff z) 
  & \propto \refMeasX(\diff z) \filt_1(z) \init_\myBlock(z_\myBlock) \obs_{1,\myBlock}(z_\myBlock) / \filt_{1,\myBlock}(z_\myBlock)\\
  & \propto \refMeasX(\diff z) \filt_1(z)  \propto \Filt_1(\diff z),
 \end{align}
 since $\smash{\filt_{1,\myBlock}(z_\myBlock) \propto \init_\myBlock(z_\myBlock) \obs_{1,\myBlock}(z_\myBlock)}$. Assume now that the statement holds at some time $t-1$. Then
 \begin{align}
  \!\!\FiltHat_{t}(\diff z)
  & \propto \int_{\spaceX} \transHat_t(x, z) \obsHat(z, y_t) \refMeasX(\diff z) \Filt_{t-1}(\diff x)\\*
  & = \frac{ \int_\spaceX \trans_\myBlock(x_{\neigh(\myBlock)}, z_\myBlock) \obs_{t,\myBlock}(z_\myBlock)\Filt_{t-1}(\diff x) }{\filt_{t,\myBlock}(z_\myBlock)}  \Filt_t(\diff z)\!\!\!\!\!\\*
  & \propto \Filt_t(\diff z),
 \end{align}
 since $\smash{\filt_{t,\myBlock}(z_\myBlock) \propto \int_\spaceX \trans_\myBlock(x_{\neigh(\myBlock)}, z_\myBlock) \obs_{t,\myBlock}(z_\myBlock) \Filt_{t-1}(\diff x)}$.
 
 Part~\ref{lem:properties_of_modelHat:2} is also proved by induction. By Part~\ref{lem:properties_of_modelHat:1}, since $\Filt_1 = \smooth_1$ and $\FiltHat_1 = \smoothHat_1$, the statement holds at time $t=1$. Assume now that the statement holds at some time~$t-1$. Then for any  $A \subseteq \spaceX_\myBlock^t$, by Part~\ref{lem:properties_of_modelHat:1},
 \begin{align}
  \!\!\MoveEqLeft\int_{\spaceX^{\mathrlap{t}}} \ind_A(x_{1:t, \myBlock}) \smoothHat_t(\diff x_{1:t})\\*
  & \propto \int_{\spaceX^{\mathrlap{t}}} \ind_A(x_{1:t, \myBlock}) \frac{\filt_{t}(x_{t})}{\filt_{t,\myBlock}(x_{t,\myBlock})} \trans_\myBlock(x_{t-1,\neigh(\myBlock)}, x_{t,\myBlock})\\*
  & \quad \times \obs_{t,\myBlock}(x_{t,\myBlock}) \smooth_{t-1}(\diff x_{1:t-1})    \refMeasX(\diff x_t)\\*
  & \propto \int_{\spaceX^{\mathrlap{t}}} \ind_A(x_{1:t, \myBlock}) \smooth_t(\diff x_{1:t}).
 \end{align}
 
 To prove Part~\ref{lem:properties_of_modelHat:3}, note that for any  $A \subseteq \spaceX_{\neigh(\myBlock)}$, by Part~\ref{lem:properties_of_modelHat:1},
 \begin{align}
  \MoveEqLeft \int_\spaceX \ind_A(z_{\neigh(\myBlock)}) \backHat_{t, \FiltHat_t}(x, \diff z)\\*
  & = \frac{\int_A \trans_\myBlock(z_{\neigh(\myBlock)}, x_\myBlock) \Filt_{t,\neigh(\myBlock)}(\diff z_{\neigh(\myBlock)})}{\int_{\spaceX_{\neigh(\myBlock)}}\trans_\myBlock(u_{\neigh(\myBlock)}, x_\myBlock) \Filt_{t,\neigh(\myBlock)}(\diff u_{\neigh(\myBlock)})}\\*
  & = \int_A \back_{\myBlock, \Filt_{t,\neigh(\myBlock)}}(x_{\myBlock}, \diff z_{\neigh(\myBlock)}).
 \end{align}
 Noting that the above expression is constant in $x_{\myBlock^\compl}$ then completes the proof of Part~\ref{lem:properties_of_modelHat:3}. \qedhere
 \end{proof}
 
 \begin{proof}[of Lemma~\ref{lem:properties_of_modelBar}]
  The proof is immediate if we replace the quantities
  $(\initHat, \transHat_t, \obsHat_t, \smoothHat_t, \FiltHat_t, \backHat_{t, \FiltHat_t}, \Filt_t, \smooth_t)$ in the proof of Lemma~\ref{lem:properties_of_modelHat} by the quantities $(\initBar, \transBar_t, \obsBar_t, \smoothBar_t, \FiltBar_t, \backBar_{t, \FiltBar_t}, \FiltTilde_t, \smoothTilde_t)$.  \qedhere
 \end{proof}

\subsection{Sufficient Statistics}

\label{subsec:sufficient_statistics}

In this subsection, we derive the sufficient statistics needed for performing parameter estimation in the high-dimensional linear-Gaussian state-space model from Section~\ref{sec:simulations} of the main document. For $q,r \in \{0, 1, \dotsc, \neighSize\}$, define 
\begin{align}
\testFun_{1,v}^{\theta, (1,r,q)} = \testFun_{1,v}^{\theta, (2,r)} \equiv 0,
\end{align}
and, for any $t \in \naturals$,
\begin{align}
 \!\testFun_{t+1,v}^{\theta, (1,r,q)}(x_{t,\neigh(v)}, x_{t+1,v}) 
 &\coloneqq \textstyle \sum_{u \in \calB_q(v)} x_{t,u} \sum_{w \in \calB_r(v)} x_{t,w},\!\!\\
 \testFun_{t+1,v}^{\theta, (2,r)}(x_{t,\neigh(v)}, x_{t+1,v}) 
 &\coloneqq x_{t+1,v} \textstyle \sum_{u \in \calB_r(v)} x_{t,u},\\
 \testFun_{t,v}^{\theta, (3)}(x_{t-1,\neigh(v)}, x_{t,v}) 
 &\coloneqq x_{t,v}^2,\\
 \testFun_{t,v}^{\theta, (4)}(x_{t-1,\neigh(v)}, x_{t,v}) 
 &\coloneqq x_{t,v} y_{t,v}. \label{eq:ssm_sufficient_statistics}
\end{align}
For any superscript $(\star)$ in the previous equation, we may then define the following sums of smoothed sufficient statistics
\begin{align}
 \smoothFunIntegral_{\nTimeSteps}^{\theta, (\star)} 
 & \coloneqq \sum_{t = 1}^\nTimeSteps \sum_{v \in \vertexSet} \E\bigl[\testFun_{t,v}^{\theta, (\star)}(X_{t-1,\neigh(v)}, X_{t,v})\bigr], \!\!\!\quad \text{$X_{1:\nTimeSteps} \sim \smooth_\nTimeSteps^\theta$.}\!\!
\end{align}
To simplify the notation, we define the following matrix and column vector 
\begin{align}
\smoothFunIntegral_{\nTimeSteps}^{\theta, (1)} & \coloneqq (\smoothFunIntegral_{\nTimeSteps}^{\theta, (1,r,q)})_{(r,q)\in \{0,\dotsc,\neighSize\}^2} \in \reals^{(\neighSize+1) \times (\neighSize+1)},\\
 \smoothFunIntegral_{\nTimeSteps}^{\theta, (2)} &\coloneqq (\smoothFunIntegral_{\nTimeSteps}^{\theta, (2,l)})_{l\in \{0,\dotsc,\neighSize\}} \in \reals^{\neighSize+1},
\end{align}
and finally, we collect all of these smoothed sufficient statistics in the ordered set
\begin{equation}
 \smoothFunIntegral_\nTimeSteps^\theta \coloneqq (\smoothFunIntegral_\nTimeSteps^{\theta, (1)}, \dotsc, \smoothFunIntegral_\nTimeSteps^{\theta, (4)}). \label{eq:smoothed_sufficient_statistics_ordered_set}
\end{equation}

\subsection{Parameter Estimation}

In this subsection, we state the ways in which the update rules of (stochastic) gradient-ascent and \gls{EM} algorithms depend on the smoothed sufficient statistics from \eqref{eq:smoothed_sufficient_statistics_ordered_set}. To simplify the notation, we write $\smash{y^2 \coloneqq \sum_{t = 1}^\nTimeSteps \textstyle\sum_{v \in \vertexSet}y_{t,v}^2}$. 

\subsubsection{Gradient-ascent Algorithm} 
While the score may be directly written as an additive function as shown in Example~\ref{ex:score} of the main document, we can alternatively write it as $\smash{\nabla_\vartheta \log \uSmooth_\nTimeSteps^\vartheta(\unitFun)|_{\vartheta = \theta} \eqqcolon \varPsi^\theta(\smoothFunIntegral_\nTimeSteps^\theta)}$. Here, for $\nTimeSteps > 1$,  writing
\begin{equation}
 \varPsi^\theta(\smoothFunIntegral_\nTimeSteps^\theta) = (\varPsi_r^\theta(\smoothFunIntegral_\nTimeSteps^\theta))_{r \in \{0,\dotsc,\neighSize+2\}},
\end{equation}
we have 
\begin{align}
 \longNegativeHSpace\varPsi_{0:\neighSize}^\theta(\smoothFunIntegral_\nTimeSteps^\theta)
 & = \eul^{-2 \theta_{\neighSize+1}} \bigl(\smoothFunIntegral_{\nTimeSteps}^{\theta,(2)} - \smoothFunIntegral_{\nTimeSteps}^{\theta,(1)}\theta_{0:\neighSize}\bigr),\\
 \longNegativeHSpace\varPsi_{\neighSize+1}^\theta(\smoothFunIntegral_\nTimeSteps^\theta)
 & = \eul^{-2 \theta_{\neighSize+1}} \bigl(\smoothFunIntegral_{\nTimeSteps}^{\theta,(3)} - \smoothFunIntegral_{1}^{\theta,(3)} - 2 \theta_{0:\neighSize}^\T \smoothFunIntegral_{\nTimeSteps}^{\theta,(2)}\\*
 & \qquad\qquad\quad  + \theta_{0:\neighSize}^\T \smoothFunIntegral_{\nTimeSteps}^{\theta,(1)} \theta_{0:\neighSize} \bigr) - \modelDim (\nTimeSteps-1),\!\!\!\!\!\!\!\\
 \longNegativeHSpace\varPsi_{\neighSize+2}^\theta(\smoothFunIntegral_\nTimeSteps^\theta)
 & =\eul^{-2 \theta_{\neighSize+2}} \bigl(\smoothFunIntegral_{\nTimeSteps}^{\theta,(3)} - 2 \smoothFunIntegral_{\nTimeSteps}^{\theta,(4)} + y^2\bigr) - \modelDim \nTimeSteps.
\end{align}

\subsubsection{\gls{EM} Algorithm} For (stochastic) \gls{EM} algorithms, the vector (of length $\neighSize+3$) needed for the update rule in \eqref{eq:offline_em} in the main document,
\begin{equation}
  \varLambda(\smoothFunIntegral_\nTimeSteps^\theta) \coloneqq (\varLambda_r(\smoothFunIntegral_\nTimeSteps^\theta))_{r \in \{0,\dotsc,\neighSize+2\}},
\end{equation}
is given by
\begin{align}
 \longNegativeHSpace\varLambda_{0:\neighSize}(\smoothFunIntegral_\nTimeSteps^\theta) & = (\smoothFunIntegral_\nTimeSteps^{\theta,(1)})^{-1}\smoothFunIntegral_\nTimeSteps^{\theta,(2)},\\
 \longNegativeHSpace\varLambda_{\neighSize+1}(\smoothFunIntegral_\nTimeSteps^\theta) 
  & = \tfrac{1}{2}\log\bigl( \tfrac{1}{\modelDim(\nTimeSteps-1)} \bigl[\smoothFunIntegral_\nTimeSteps^{\theta,(3)} - \smoothFunIntegral_1^{\theta,(3)}\\*
  & \qquad\qquad - (\smoothFunIntegral_\nTimeSteps^{\theta,(2)})^\T (\smoothFunIntegral_\nTimeSteps^{\theta,(1)})^{-1} \smoothFunIntegral_\nTimeSteps^{\theta,(2)}\bigr]\bigr),\\
 \longNegativeHSpace\varLambda_{\neighSize+2}(\smoothFunIntegral_\nTimeSteps^\theta)
  & = \tfrac{1}{2}\log\bigl(\tfrac{1}{\modelDim\nTimeSteps}\bigl[\smoothFunIntegral_\nTimeSteps^{\theta,(3)} - 2 \smoothFunIntegral_\nTimeSteps^{\theta,(4)} + y^2\bigr).
\end{align}
Finally, we note that since this model is in the exponential family, the maximisation problem admits a closed-form solution.

\subsection{Implementation Details}

All the computations in this paper were implemented in the programming language C++ using the Armadillo linear algebra library \citep{sanderson2016armadillo}. All the algorithms were called from the R programming language \citep{r2015r} using various Rcpp \citep{eddelbuettel2011rcpp, eddelbuettel2013seamless} libraries.

\end{document}

%% file: block_abbreviations.tex
\newacronym{ABC}{ABC}{approximate Bayesian computation}%
\newacronym{PMCMC}{PMCMC}{particle Markov chain Monte Carlo}%
\newacronym[\glslongpluralkey={(general state-space) hidden Markov models}]{HMM}{HMM}{(general state-space) hidden Markov model}%
\newacronym{ESS}{ESS}{effective sample size}%
\newacronym{CESS}{CESS}{conditional effective sample size}%
\newacronym{RM}{RM}{resample--move}%
\newacronym{PDF}{PDF}{probability density function}%
\newacronym{IID}{IID}{independent and identically distributed}%
\newacronym{MCMC}{MCMC}{Markov chain Monte Carlo}%
\newacronym{MH}{MH}{Metropolis--Hastings}%
\newacronym{SA}{SA}{simulated annealing}%
\newacronym{PMMH}{PMMH}{particle marginal Metro\-po\-lis--Has\-tings}%
\newacronym{GIMH}{GIMH}{grouped independent Metropolis--Hastings}%
\newacronym{MCWM}{MCWM}{Monte Carlo within Metropolis}%
\newacronym{CDF}{CDF}{cumulative distribution function}%
\newacronym{SMC}{SMC}{sequential Monte Carlo}%
\newacronym{PPP}{ppp}{Poisson point process}%
\newacronym{CSMC}{CSMC}{conditional sequential Monte Carlo}%
\newacronym{SLLN}{SLLN}{strong law of large numbers}%
\newacronym{EM}{EM}{expectation--maximisation}%
\newacronym[user1={importance-sampling}]{IS}{IS}{importance sampling}%
\newacronym{SNIS}{SNIS}{self-normalised importance sampling}%
\newacronym{WLLN}{WLLN}{weak law of large numbers}%
\newacronym{CLT}{CLT}{central limit theorem}%
\newacronym{MSE}{MSE}{mean-square error}%
\newacronym{RMSE}{RMSE}{root-mean-square error}%
\newacronym{RJMCMC}{RJMCMC}{reversible-jump Markov chain Monte Carlo}%
\newacronym{ML}{ML}{maximum likelihood}%
\newacronym{MLE}{MLE}{maximum-likelihood estimate}%
\newacronym{FFBS}{FFBS}{forward filtering--backward sampling}%
\newacronym[user1={ancestor-sampling}]{AS}{AS}{ancestor sampling}%
\newacronym[user1={backward-sampling}]{BS}{BS}{backward sampling}%
\newacronym[user1={forwarwd-smoothing}]{FS}{FS}{forward smoothing}%
\newacronym{BPF}{BPF}{blocked particle filter}%
\newacronym{PF}{PF}{particle filter}%
\newacronym{FMRI}{FMRI}{functional magnetic resonance imaging}%
\newacronym{as}{a.s.}{almost surely}%
\newacronym{PaRIS}{PaRIS}{particle-based rapid incremental smoother}%
\newacronym{SV}{SV}{stochastic volatility}

%% file: block_mathematics.tex
\DeclareMathOperator*{\argmax}{arg\,max}%
\DeclareMathOperator{\Prob}{\mathbb{P}}%
\DeclareMathOperator{\E}{\mathbb{E}}%
\DeclareMathOperator{\MSE}{\mathit{MSE}}%
\DeclareMathOperator{\ind}{\bm{\mathrm{1}}}%
\newcommand{\unitFun}{\mathbf{1}}%
\DeclareMathOperator{\dN}{N}%
\DeclareMathOperator{\dCat}{Cat}%

\newcommand{\mathsc}[1]{{\normalfont\textsc{#1}}}%

\newcommand{\testFun}{f}%
\newcommand{\reals}{\mathbb{R}}%
\newcommand{\naturals}{\mathbb{N}}%

\DeclareMathOperator{\Borel}{\mathrm{Borel}}%

\DeclareMathOperator{\bo}{\mathcal{O}}

\newcommand{\coloneqq}{\mathrel{\vcentcolon=}}%
\newcommand{\eqqcolon}{\mathrel{=\vcentcolon}}%

\newcommand{\ccdot}{\,\cdot\,}
\newcommand{\diff}{\mathrm{d}}
\newcommand{\T}{\mathsf{T}}
\newcommand{\compl}{\mathsc{c}}
\newcommand{\eul}{\mathrm{e}}%

\makeatletter
\let\save@mathaccent\mathaccent
\newcommand*\if@single[3]{%
  \setbox0\hbox{${\mathaccent"0362{#1}}^H$}%
  \setbox2\hbox{${\mathaccent"0362{\kern0pt#1}}^H$}%
  \ifdim\ht0=\ht2 #3\else #2\fi
  }
\newcommand*\rel@kern[1]{\kern#1\dimexpr\macc@kerna}
\newcommand*\widebar[1]{\@ifnextchar^{{\wide@bar{#1}{0}}}{\wide@bar{#1}{1}}}
\newcommand*\wide@bar[2]{\if@single{#1}{\wide@bar@{#1}{#2}{1}}{\wide@bar@{#1}{#2}{2}}}
\newcommand*\wide@bar@[3]{%
  \begingroup
  \def\mathaccent##1##2{%
    \let\mathaccent\save@mathaccent
    \if#32 \let\macc@nucleus\first@char \fi
    \setbox\z@\hbox{$\macc@style{\macc@nucleus}_{}$}%
    \setbox\tw@\hbox{$\macc@style{\macc@nucleus}{}_{}$}%
    \dimen@\wd\tw@
    \advance\dimen@-\wd\z@
    \divide\dimen@ 3
    \@tempdima\wd\tw@
    \advance\@tempdima-\scriptspace
    \divide\@tempdima 10
    \advance\dimen@-\@tempdima
    \ifdim\dimen@>\z@ \dimen@0pt\fi
    \rel@kern{0.6}\kern-\dimen@
    \if#31
      \overline{\rel@kern{-0.6}\kern\dimen@\macc@nucleus\rel@kern{0.4}\kern\dimen@}%
      \advance\dimen@0.4\dimexpr\macc@kerna
      \let\final@kern#2%
      \ifdim\dimen@<\z@ \let\final@kern1\fi
      \if\final@kern1 \kern-\dimen@\fi
    \else
      \overline{\rel@kern{-0.6}\kern\dimen@#1}%
    \fi
  }%
  \macc@depth\@ne
  \let\math@bgroup\@empty \let\math@egroup\macc@set@skewchar
  \mathsurround\z@ \frozen@everymath{\mathgroup\macc@group\relax}%
  \macc@set@skewchar\relax
  \let\mathaccentV\macc@nested@a
  \if#31
    \macc@nested@a\relax111{#1}%
  \else
    \def\gobble@till@marker##1\endmarker{}%
    \futurelet\first@char\gobble@till@marker#1\endmarker
    \ifcat\noexpand\first@char A\else
      \def\first@char{}%
    \fi
    \macc@nested@a\relax111{\first@char}%
  \fi
  \endgroup
}
\makeatother

\newcommand{\calB}{\mathcal{B}}%
\newcommand{\calS}{\mathcal{S}}%

\newcommand{\longNegativeHSpace}{\!\!\!\!\!\!\!\!\!\!\!\!}

\newcommand*{\mycmss}{\fontfamily{cmss}\selectfont}
\DeclareTextFontCommand{\textcmss}{\mycmss}

\newcommand{\Filt}{\pi}%
\newcommand{\filt}{\varpi}%
\newcommand{\uSmooth}{\varGamma}%
\newcommand{\smooth}{\mathbb{Q}}%
\newcommand{\trans}{p}%
\newcommand{\obs}{g}%
\newcommand{\prop}{q}%
\newcommand{\Prop}{Q}%
\newcommand{\init}{m}%
\newcommand{\Init}{\mu}%
\newcommand{\back}{B}%
\newcommand{\myBlockSet}{\mathcal{K}}%
\newcommand{\vertexSet}{\mathbb{V}}%
\newcommand{\myBlock}{K}%
\newcommand{\enlargedBlock}{{\smash{\widebar{K}}}}%
\newcommand{\edgeSet}{\mathcal{E}}%
\newcommand{\graphSet}{\mathcal{G}}%
\newcommand{\neigh}{\mathcal{N}}
\newcommand{\spaceX}{\mathbb{X}}%
\newcommand{\spaceY}{\mathbb{Y}}%
\newcommand{\refMeasX}{\psi}%
\newcommand{\refMeasY}{\varphi}%
\DeclareMathOperator{\card}{\mathrm{card}}
\DeclareMathOperator{\metric}{\mathit{d}}%
\DeclareMathOperator{\osc}{osc}%
\DeclareMathOperator{\Osc}{Osc}%
\newcommand{\modelDim}{V}%
\newcommand{\nTimeSteps}{T}%
\newcommand{\nParticles}{N}%
\newcommand{\nBackwardPaths}{M}%
\newcommand{\myBlockSetSize}{\card \mathcal{K}}%
\newcommand{\maxBlockSize}{\lvert \myBlockSet \rvert_\infty}%
\newcommand{\neighSize}{R}%
\newcommand{\smoothFun}{F}%
\newcommand{\smoothFunIntegral}{\mathbb{\smoothFun}}%

\newcommand{\smoothFunComp}{\mathring{F}}%
\newcommand{\FiltComp}{\mathring{\Filt}}%
\newcommand{\spaceXComp}{\mathring{\spaceX}}%
\newcommand{\spaceYComp}{\mathring{\spaceY}}%
\newcommand{\transComp}{\mathring{p}}%
\newcommand{\obsComp}{\mathring{g}}%
\newcommand{\QComp}{\mathring{Q}}%
\newcommand{\QCalComp}{\mathring{\mathcal{Q}}}%
\newcommand{\PComp}{\mathring{P}}%
\newcommand{\DComp}{\mathring{D}}%
\newcommand{\backComp}{\mathring{\back}}%
\newcommand{\GComp}{\mathring{G}}%
\newcommand{\smoothComp}{\mathring{\smooth}}%
\newcommand{\refMeasXComp}{\mathring{\refMeasX}}%
\newcommand{\refMeasYComp}{\mathring{\refMeasY}}%

\newcommand{\pf}{pf}%

\newcommand{\bpf}{bpf}%

\newcommand{\bs}{\mathsc{bs}}%

\newcommand{\fs}{\mathsc{fs}}%
\newcommand{\fsBlock}[1]{\mathsc{fs}({#1})}

\newcommand{\transBar}{\bar{\trans}}%
\newcommand{\obsBar}{\bar{\obs}}%
\newcommand{\FiltBar}{\bar{\pi}}%
\newcommand{\initBar}{\widebar{\init}}%
\newcommand{\smoothBar}{\widebar{\smooth}}%
\newcommand{\backBar}{\widebar{B}}%

\newcommand{\transHat}{\hat{\trans}}%
\newcommand{\obsHat}{\hat{\obs}}%
\newcommand{\FiltHat}{\hat{\pi}}%
\newcommand{\initHat}{\widehat{\init}}%
\newcommand{\smoothHat}{\widehat{\smooth}}%
\newcommand{\backHat}{\widehat{B}}%

\newcommand{\transTilde}{\tilde{\trans}}%
\newcommand{\obsTilde}{\tilde{\obs}}%
\newcommand{\FiltTilde}{\tilde{\pi}}%
\newcommand{\filtTilde}{\widetilde{\varpi}}%
\newcommand{\initTilde}{\widetilde{\init}}%
\newcommand{\smoothTilde}{\widetilde{\smooth}}%

\newcommand{\modelOriginal}{(\init, \trans, \obs_t)}%
\newcommand{\modelTilde}{(\initTilde, \transTilde_t, \obsTilde_t)}%
\newcommand{\modelBar}{(\initBar, \transBar_t, \obsBar_t)}%
\newcommand{\modelHat}{(\initHat, \transHat_t, \obsHat_t)}%

\newcommand{\consta}{c_1}%
\newcommand{\constb}{c_2}%
\newcommand{\constc}{c_3}%

\newcommand{\thetaEstimate}{\theta}%
\newcommand{\qedhere}{\hfill \ensuremath{_\square}}